\documentclass[aps,floatfix,showpacs,superscriptaddress,showkeys]{revtex4} 

\usepackage{amssymb,amsmath,amsfonts,latexsym,graphicx,epsfig,here}

\usepackage{mkPhys}

\newcommand{\bea}{\begin{eqnarray}}
\newcommand{\ena}{\end{eqnarray}} 

\newcommand{\be}{\begin{equation}}
\newcommand{\en}{\end{equation}}

\newcommand{\nn}{\nonumber\\}

\begin{document}

\vspace*{0.2cm}
\hfill DSF-2011-14 (Napoli), MZ-TH/11-44 (Mainz)
\vspace*{0.2cm}

\title{Form factors for semileptonic, nonleptonic and rare 
\boldmath{$B\,(B_s)$} meson decays}

\author{Mikhail A. Ivanov}
\affiliation{Bogoliubov Laboratory of Theoretical Physics, \\
Joint Institute for Nuclear Research, 141980 Dubna, Russia}

\author{J\"{u}rgen G. K\"{o}rner}
\affiliation{Institut f\"{u}r Physik, Johannes Gutenberg-Universit\"{a}t, \\
D-55099 Mainz, Germany}

\author{Sergey G. Kovalenko}
\affiliation{Centro de Estudios Subat\'omicos (CES),
Universidad T\'ecnica Federico Santa Mar\'\i a, \\
Casilla 110-V, Valpara\'\i so, Chile}

\author{Pietro Santorelli}
\affiliation{Dipartimento di Scienze Fisiche, Universit\`a di Napoli
Federico II, Complesso Universitario di Monte S. Angelo,
Via Cintia, Edificio 6, 80126 Napoli, Italy, and
Istituto Nazionale di Fisica Nucleare, Sezione di Napoli}

\author{Gozyal G. Saidullaeva}
\affiliation{Al-Farabi Kazak National University, 480012 Almaty, Kazakhstan}

\begin{abstract}
We provide new values for the model parameters of the covariant constituent 
quark model (with built--in infrared confinement) in the meson sector by 
a fit to the leptonic decay constants and a number of
electromagnetic decays. We then evaluate, in a parameter-free way, the form 
factors of the $B(B_s)\to P(V)$ transitions in the full kinematical
region of momentum transfer.  
As an application of our results we calculate
the widths of the nonleptonic $B_s$-decays into 
$D_s^- D_s^+,$  $D_s^{\ast\,-} D_s^{+}+D_s^- D_s^{\ast\,+}$ and
$D_s^{\ast\,-} D_s^{\ast\,+}$. These modes give the largest contribution
to $\Delta\Gamma$ for the $B_s-\bar B_s$ system.
We also treat the nonleptonic decay $B_{s}\to\Jpsi\phi$. Although this mode is 
color-suppressed, this decay has important implications for the search of 
possible CP-violating new physics effects in $B_s-\bar B_s$ mixing.

\end{abstract}

\maketitle

\section{Introduction}
\label{sec:intro}

The study of heavy flavor physics is important due to the unique
possibility of determining the Cabibbo-Kobayashi-Maskawa
matrix elements. Such studies also provide insights into the origin
of flavor and CP-violation. Moreover, one of the main
purposes of heavy-flavor experiments is to look for
new physics beyond the standard model (see the recent
review~\cite{Stone}). The subject to study are
heavy hadrons containing a $b$-- or a $c$--quark and
their weak decays. Note that the $t$ quark decays too quickly
to form stable hadrons. Recently, time--dependent measurements of CP 
violation in the $B_s-\bar B_s$ system have become available. In the wake
of these measurements, the decay
$B_s\to \Jpsi\,\phi$ has attracted much attention from
both theorists and experimentalists (see, for instance,
Refs.~\cite{leitner11,Faller:2008gt} and references therein).

The main idea in the theoretical studies of heavy-flavor decays
is to separate short-distance (perturbative) QCD dynamics
from long-distance (nonperturbative) hadronic effects.
One uses the so-called {\it naive} factorization approach
which is based on the weak effective Hamiltonian describing
quark and lepton transitions in terms of local operators
that are multiplied by Wilson coefficients
(for a review, see Ref.~\cite{Buchalla:1995vs}).
The Wilson coefficients characterize the short-distance
dynamics and may be reliably evaluated by 
perturbative methods. The calculation of the hadronic matrix
elements of local operators between initial and final states
require nonperturbative methods. One needs to know how
hadrons are constructed from quarks. Technically,
any matrix element of a local operator may be expressed
in terms of a set of scalar functions which are referred to as form factors.
The so-called QCD factorization and the soft-collinear effective theory 
yield factorization theorems which allow for a systematic
description of a given process in terms
of products of soft and hard matrix elements
(we refer an interested reader to Refs.~\cite{Beneke:1999br}, 
\cite{Feldmann:2008sk} and \cite{Bauer:2000yr}).

A variety of theoretical approaches have been used to evaluate
the hadronic form factors. The least model dependent among these is the
light-cone sum rule (LCSR) approach  (see Refs.~\cite{Ball} and
\cite{Khodjamirian:2006st}). In the LCRS approach one can 
access the form factors in the large recoil (small momentum transfer) 
region which are then extrapolated to the near-zero recoil region
using some model-dependent pole--type parametrizations. 
Grinstein and Pirjol have developed a systematic approach to the rare decay 
$B\to K^\ast \ell^+\ell^-$ in the low recoil region using the heavy-quark 
effective theory framework~\cite{Grinstein:2004vb}. The low-recoil approach 
was late studied in detail in Ref.~\cite{Hiller}.

We mention a few other model approaches for the calculation of the form 
factors. They are based on the study of i) Dyson-Schwinger equations in QCD
\cite{Ivanov:2007cw}; ii) the constituent quark model using 
dispersion relations \cite{Melikhov}; iii) a relativistic quark model
developed by
Ebert, Faustov and Galkin \cite{Ebert:2006nz};
iv) a QCD relativistic potential model \cite{Ladisa:1999cu} (see also \cite{hep-ph/9809372});
v) a QCD sum rule analysis \cite{Colangelo:1995jv};
and, finally, vi) the covariant constituent quark model developed by some of us
starting with Refs.~\cite{Faessler:2002ut}, \cite{IKS} and 
\cite{Ivanov:2002un}.
It is worth mentioning that the entire physical range of momentum transfer is 
accessible in the covariant quark model approach used in  
Refs.~\cite{Faessler:2002ut,IKS,Ivanov:2002un} and the present paper and 
in the calculations of Refs.~\cite{Ivanov:2007cw,Melikhov,Ebert:2006nz,
Ladisa:1999cu,Colangelo:1995jv}.

The earlier versions of the covariant constituent quark model (for short:
covariant quark model \footnote{Previously we referred to the covariant quark 
model as the relativistic quark model. We decided to rechristen its name
in order to set it apart from other relativistic
quark models as e.g. the one developed by
Ebert, Faustov and Galkin \cite{Ebert:2006nz} which contain relativistic 
elements but are not truly covariant.}) in Refs.\,\cite{Faessler:2002ut,IKS,Ivanov:2002un} 
did not include the confinement of quarks. The nonconfined version 
had been applied, among others, to the description of $B$ and $B_c$ transition 
form factors using a small set of model parameters. In the covariant
quark model, meson transitions are described by covariant Feynman
diagrams with free constituent quark propagation. The ultraviolet 
behavior of the loop diagrams is tempered by appropriately damped vertex 
functions. A key role in the consistent formulation of the model is played 
by the so-called compositeness condition \cite{Z=0,Z=0_2}, a corollary of 
which guarantees the correct charge normalization of charged mesons at zero
momentum transfer. Since the propagation of the constituent quarks is 
described by free-particle Green's functions, one will encounter on--shell 
quark production in the case when the mass of the bound state
exceeds the sum of the constituent quark masses. Therefore, the
applicability of the covariant quark model in its original version
was limited to the cases where $m_H<m_{q_1}+m_{q_2}$. This limitation 
was removed later on in Ref.~\cite{Branz:2009cd} by effectively introducing 
infrared confinement through the introduction of a universal infrared
cutoff parameter in the space of loop integrations. This extended the
applicability of the covariant quark model to all processes involving
heavy and light hadrons. The viability of the improved covariant quark model
was demonstrated in a number of applications to mesonic transitions
in Ref.~\cite{Branz:2009cd}. Later on, this approach was successfully applied to 
a study of the tetraquark state X(3872) and its strong and radiative decays
(see Refs.~\cite{Dubnicka:2010kz,Dubnicka:2011mm}).

Once the parameters of the covariant quark model have been determined the 
covariant quark model is a very flexible tool that can be used to calculate 
any heavy--to--heavy,
heavy--to--light and light--to--light hadron transition. While the more model
independent approaches usually have to rely on a heavy quark mass expansion 
the predictions of the covariant quark model hold for general mass
configurations which are not accessible to the model-independent approaches. 
On the other hand the predictions of the heavy-quark expansion can be 
recovered by using static propagators for the heavy quarks.

In this paper we use the improved version of the covariant quark model 
including infrared confinement to evaluate the form factors
of the $B(B_s)\to P(V)$ transitions in the full kinematical
range of momentum transfer.  
As an application of our results we calculate
the widths of several $B_s$ nonleptonic decays. 
These are the modes  
$B_{s} \to D_s^- D_s^+,  D_s^{\ast\,-} D_s^{+}+D_s^- D_s^{\ast\,+}$, and
$B_{s} \to D_s^{\ast\,-} D_s^{\ast\,+}$ which give the largest contribution
to $\Delta\Gamma$ for the $B_s-\bar B_s$ system. We also treat the
color-suppressed mode $B_{s} \to \Jpsi\phi$. This decay
is important for the search of possible CP-violating 
new physics effects in $B_s-\bar B_s$ mixing.

Our paper is structured as follows.
In Sec.~\ref{sec:model}, we give a brief sketch of the
theoretical framework underlying the covariant quark model including a 
discussion of how infrared confinement is set up in the loop integrations.
In Sec.~\ref{sec:modelpar}, we discuss in some detail how the
model parameters of the covariant quark model are determined
through a least-squares fit to experimental/theoretical data on leptonic
decay constants and to eight fundamental mesonic one-- and two--photon 
decays. Once the model
parameters of the covariant quark model are fixed, the model can be used to 
obtain parameter-free predictions for any transition process involving light 
or heavy mesons. In Sec.~\ref{sec:tff}, we calculate the transition form 
factors of the $B$ and $B_{s}$ mesons to light pseudoscalar and vector mesons
which are needed as ingredients for the calculation of the semileptonic,
nonleptonic and 
rare decays of the $B$ and $B_{s}$ mesons.
In Sec.~\ref{sec:nonlep}, we make use of the calculated form factors to
calculate the nonleptonic decays $B_{s} \to D_{s}^{(\ast)-} D_{s}^{(\ast)+}$
and $B_{s} \to \Jpsi\phi$ which have recently attracted some attention as
explained above. Finally, in Sec.~\ref{sec:summary}, we summarize our findings.

\section{Effective Lagrangians,  Compositeness Condition, and
Infrared Confinement}
\label{sec:model}
In this section, we give a brief description of the theoretical framework
underlying the formulation of the covariant quark model. We first define a 
nonlocal meson-quark-quark vertex in terms of an effective 
Lagrangian. We then introduce the compositeness condition and discuss its
significance. We then, finally, describe how infrared confinement is
incorporated into the model. This involves a technical discussion of how the 
one--loop integrations are done, which we briefly describe.  

The coupling of a meson $H(q_1\bar q_2)$ to its constituent
quarks $q_1$ and $\bar q_2 $ is described  by the effective Lagrangian 
\cite{QCM,RCQM}
\be
\label{eq:Lagr}
{\mathcal L}_{\rm int\, Hqq}(x) = g_H H(x)\int\!\! dx_1 \!\!\int\!\!
dx_2 F_H (x,x_1,x_2)\bar q_2(x_2)\Gamma_H q_1(x_1) \, + {\rm h.c.}
\en
$\Gamma_H$ is a Dirac matrix or a string of Dirac matrices
which projects onto the spin quantum number of the meson field $H(x)$.
In the present case, the Dirac structures involved are $\gamma_{5}$ for the
pseeudoscalar meson and $\gamma_{\mu}$ for the vector meson.
The function $F_H$ is related to the scalar part of the
Bethe-Salpeter amplitude and characterizes the finite size of the
meson. To satisfy translational invariance, the scalar function $F_H$
has to fulfill the relation $F_H(x+a,x_1+a,x_2+a)=F_H(x,x_1,x_2)$ for
any four-vector a. 
A specific form which satisfies translational invariance is the form
\be
\label{eq:vertex}
F_H(x,x_1,x_2)=\delta(x - w_1 x_1 - w_2 x_2) \Phi_H((x_1-x_2)^2)
\en
where $\Phi_H$ is the correlation function of the two constituent quarks
with masses $m_{q_1}$, $m_{q_2}$, and the mass ratios
$w_i = m_{q_i}/(m_{q_1}+m_{q_2})$.

The coupling constant $g_H$ in Eq.~(\ref{eq:Lagr}) is constrained by the
so-called {\it compositeness condition} originally proposed 
in Refs.~\cite{Z=0,Z=0_2},
and extensively used in Refs.\,\cite{QCM,RCQM}.
The compositeness condition requires that the renormalization constant of
the elementary meson field $H(x)$ is set to zero:
\be
\label{eq:Z=0}
Z_H \, = \, 1 - \, \frac{3g^2_H}{4\pi^2} \,
\widetilde\Pi'_H(m^2_H) \, = \, 0
\en
where $\widetilde\Pi^\prime_H$ is the derivative of the meson mass operator.

To clarify the physical meaning of the compositeness condition 
in Eq.~(\ref{eq:Z=0}), we first want to remind the reader
that the renormalization constant $Z_H^{1/2}$ can also interpreted as
the matrix element between the physical and the corresponding bare state.
The condition  $Z_H=0$ implies that the physical state does not contain
the bare state and is appropriately described as a bound state.
The interaction Lagrangian of Eq.~(\ref{eq:Lagr}) and
the corresponding free parts of the Lagrangian describe
both the constituents (quarks) and the physical particles (hadrons)
which are viewed as the bound states of the quarks.
As a result of the interaction, the physical particle is dressed,
i.e. its mass and wave function have to be renormalized.

In a more familiar setting the compositeness condition $Z_{H}=0$ guarantees 
the correct charge normalization of a charged particle at zero momentum 
transfer. This can be seen by using an identity relating the derivative
of the free-quark propagator (with loop momentum $k+p$) with the 
electromagnetic $\gamma_{\mu}$ coupling to the same propagator at zero momentum transfer.
The identity reads
\begin{equation}
\label{norm}
\frac{\partial}{\partial p^{\mu}}\frac{1}{m_{q}-\not\! k-\not\! p}=
\frac{1}{m_{q}-\not\! k-\not\! p}\,\gamma_{\mu}\,
\frac{1}{m_{q}-\not\! k-\not\! p}\,. 
\end{equation}
The contribution of the left-hand-side of Eq.(\ref{norm}) is normalized due to the 
compositeness condition, and, therefore, the contribution of the right-hand-side is also 
normalized.

The condition $Z_H=0$ also effectively excludes
the constituent degrees of freedom from the space of physical states.
It thereby guarantees that there is no double counting
for the physical observable under consideration.
The constituents exist only in  virtual states.
One of the corollaries of the compositeness condition is the absence
of a direct interaction of the dressed charged particle with the
electromagnetic field. Taking into account both the tree-level
diagram and the diagrams with self-energy insertions into the
external legs (i.e. the tree-level diagram times $Z_H -1$) yields
a common factor $Z_H$, which is equal to zero. We refer the interested
reader to our previous papers
\cite{QCM,RCQM,IKS}
where these points are discussed in more detail.

In the case of pseudoscalar and vector mesons,
the derivative of the meson mass operator appearing in Eq.~(\ref{eq:Z=0})
can be calculated from the one-loop two--point function given by

\bea\label{eq:Mass-operator}
\widetilde\Pi'_P(p^2) &=&
\frac{1}{2p^2}\,
p^\alpha\frac{d}{dp^\alpha}\,
\int\!\! \frac{d^4k}{4\pi^2i}\, \widetilde\Phi^2_P(-k^2)\,
{\rm tr} \biggl[\gamma^5 S_1(k+w_1 p)\gamma^5 S_2(k-w_2 p) \biggr] 
\nn
&=&
\frac{1}{2p^2}\,\int\!\! \frac{d^4k}{4\pi^2i}\,\widetilde\Phi^2_P(-k^2)\,
\Big\{
w_1\,{\rm tr} 
\biggl[\gamma^5 S_1(k+w_1 p)\!\not\!p\, S_1(k+w_1 p)\gamma^5 S_2(k-w_2 p) \biggr] 
\nn
&&
\phantom{
\frac{1}{2p^2}\,\int\!\! \frac{d^4k}{4\pi^2i}\,\widetilde\Phi^2_P(-k^2)\,
\!\!\!
}
-w_2\,{\rm tr} 
\biggl[\gamma^5 S_1(k+w_1 p)\gamma^5 S_2(k-w_2 p)\!\not\!p\, S_2(k-w_2 p) \biggr]
\Big\} 
\nn
\nn
 \tilde\Pi'_V(p^2)  &=&
\frac{1}{3}\left( g_{\mu\nu}-\frac{p_\mu p_\nu}{p^2}\right)
\frac{1}{2p^2}\,p^\alpha\frac{d}{dp^\alpha}\,
\int\!\! \frac{d^4k}{4\pi^2i}\, \widetilde\Phi^2_V(-k^2)\,
{\rm tr} \biggl[\gamma^\mu S_1(k+w_1 p) \gamma^\nu S_2(k-w_2 p) \biggr]
\nn
&=&
\frac{1}{3}\left( g_{\mu\nu}-\frac{p_\mu p_\nu}{p^2}\right)
\frac{1}{2p^2}\,\int\!\! \frac{d^4k}{4\pi^2i}\,\widetilde\Phi^2_V(-k^2)\,
\Big\{
w_1\,{\rm tr} 
\biggl[
\gamma^\mu S_1(k+w_1 p)\!\not\!p\, S_1(k+w_1 p)\gamma^\nu S_2(k-w_2 p) 
\biggr] 
\nn
&&
\phantom{
\frac{1}{3}\left( g_{\mu\nu}-\frac{p_\mu p_\nu}{p^2}\right)
\frac{1}{2p^2}\,\int\!\! \frac{d^4k}{4\pi^2i}\,\widetilde\Phi^2_V(-k^2)\,
\!\!\!
}
-w_2\,{\rm tr} 
\biggl[
\gamma^\mu S_1(k+w_1 p)\gamma^\nu S_2(k-w_2 p)\!\not\!p\, S_2(k-w_2 p) \biggr]
\Big\}, 
\ena
where $\widetilde\Phi_H(-k^2)$ is the Fourier-transform of
the vertex function $\Phi_H((x_1-x_2)^2)$, $S_i(k)$ is the free-quark propagator given by
\be\label{eq:quark-prop}
S_i(k)=\frac{1}{m_{q_i}-\not\! k}\,,
\en
and $m_{q_i}$ is the effective constituent quark mass $m_{q_i}$. 

For calculational convenience, we will choose a simple Gaussian form for the 
vertex function $\bar \Phi_H(-\,k^2)$. 
One has
\be
\bar \Phi_H(-\,k^2) 
= \exp\left(k^2/\Lambda_H^2\right)
\label{eq:Gauss}
\en
where the parameter $\Lambda_H$ characterizes the size of the respective 
bound state 
meson $H$. Since $k^2$ turns into $-\,k_E^2$ in Euclidean space, the form
(\ref{eq:Gauss}) has the appropriate fall-off behavior in the Euclidean region.
We emphasize that any choice for  $\Phi_H$ is appropriate
as long as it falls off sufficiently quickly in the ultraviolet region of
Euclidean space to render the corresponding Feynman diagrams ultraviolet 
finite. 

The technical details of how the one--loop integrations such as in
Eq.(\ref{eq:Mass-operator}) are done can be found in Ref.~\cite{Branz:2009cd}.
Let us mention that we use the Schwinger representation to write the
local quark propagators as
\be
\label{schwinger}
S(k) = (m+\not\! k)
\int\limits_0^\infty\! 
d\beta\,e^{-\beta\,(m^2-k^2)}\,.
\en
The loop momentum now appears in the exponent which allows one to deal
very efficiently with tensor loop integrals by converting loop momenta
into derivatives via the identity
\be
k_i^\mu e^{2kr} = \frac{1}{2}\frac{\partial}{\partial r_{i\,\mu}}e^{2kr},
\en 
We have written a FORM \cite{Vermaseren:2000nd} program that achieves 
the necessary commutations of the differential operators in a very efficient 
way.   

After doing the loop integration one obtains
\be
\Pi =  \int\limits_0^\infty d^n \beta \, F(\beta_1,\ldots,\beta_n) \,,
\en
for a given Feynman diagram $\Pi$,
where $F$ stands for the whole structure of a given diagram. For the
mass operators of Eq.~(\ref{eq:Mass-operator}) one has three propagators, and, thus,
one has three Schwinger parameters $\beta_{i}$ $(i=1,2,3)$. For the transition form 
factors to be discussed later on, one has again three propagators leading
again to $n=3$.   

Next, we briefly describe how infrared confinement is 
implemented~\cite{Branz:2009cd} in the quark loops.
First, note that the set of Schwinger parameters $\beta_i$ can be turned into 
a simplex by introducing an additional $t$ integration via the identity 
\be 
1 = \int\limits_0^\infty dt \, \delta(t - \sum\limits_{i=1}^n \beta_i)
\en 
leading to 
\be
\hspace*{-0.2cm}
\Pi   = \int\limits_0^\infty\! dt\, t^{n-1}\!\! \int\limits_0^1\! d^n \alpha \, 
\delta\Big(1 - \sum\limits_{i=1}^n \alpha_i \Big) \, 
F(t\alpha_1,\ldots,t\alpha_n). 
\label{eq:loop_2} 
\en
There are now altogether $n$ numerical integrations: 
$(n-1)$ $\alpha$--parameter
integrations and the integration over the scale parameter $t$. 
The very large $t$ region corresponds to the region where the singularities
of the diagram with its local quark propagators start appearing. 
However, as described in Ref.\,\cite{Branz:2009cd}, if one introduces 
an infrared cut-off on the upper limit of the t integration, all 
singularities vanish because the integral is now analytic for any value
of the kinematic variables.
We cutoff the upper integration at $1/\lambda^2$ and obtain
\be
\hspace*{-0.2cm}
  \Pi^c = \!\!  
\int\limits_0^{1/\lambda^2}\!\! dt\, t^{n-1}\!\! \int\limits_0^1\! d^n \alpha \, 
\delta\Big(1 - \sum\limits_{i=1}^n \alpha_i \Big) \, 
F(t\alpha_1,\ldots,t\alpha_n).
\en  
By introducing the infrared cutoff one has removed all potential thresholds 
in the quark-loop diagram, i.e. the quarks are never on-shell and are thus
effectively confined. We take the cutoff parameter $\lambda$ to be the 
same in all physical processes, i.e. the infrared parameter is universal.
The numerical evaluation of the integrals have been done by a numerical 
program written in FORTRAN code.

\section{Model Parameters}
\label{sec:modelpar}
Let us first enumerate the number of model parameters in the covariant quark 
model. For a given meson $H_{i}$ these are the coupling parameter 
$g_{H_{i}}$, the size parameter $\Lambda_{H_{i}}$, two of the four effective 
constituent quark masses 
$,m_{q_{j}}(m_{u}=m_{d}, m_{s}, m_{c}, m_{b})$, and the universal confinement
parameter $\lambda$. For $n_{H}$ mesons one therefore has $2n_{H}+5$ model 
parameters. The compositeness condition provides $n_{H}$ constraints on the 
model parameters, which we symbolically write as 
\be
\label{compcon}
f_{H_{i}}\,(g_{H_{i}},\Lambda_{H_{i}},m_{q_{i}},\lambda)=1
\en
The constraint (\ref{compcon}) can be used, e.g., to eliminate the coupling
parameter $g_{H}$ from the set of parameters. The remaining parameters are
determined by a fit to experimental data. An obvious choice is to fit the 
model parameters to the experimental values of the leptonic decay constants.
In the covariant quark model the relevant expressions for the pseudoscalar
and vector mesons are given by
\bea\label{eq:lept}
N_c\, g_P\! \int\!\! \frac{d^4k}{ (2\pi)^4 i}\, \widetilde\Phi_P(-k^2)\,
{\rm tr} \biggl[O_{P}^{\,\mu} S_1(k+w_1 p) \gamma^5 S_2(k-w_2 p) \biggr] 
&=&f_P p^\mu, \phantom{m_V}\qquad p^2=m^2_P,
\nn
N_c\, g_V\! \int\!\! \frac{d^4k}{ (2\pi)^4 i}\, \widetilde\Phi_V(-k^2)\,
{\rm tr} \biggl[O_{V}^{\,\mu} S_1(k+w_1 p)\not\!\epsilon_V  S_2(k-w_2 p) 
\biggr] &=& m_V f_V \epsilon_V^\mu,\qquad p^2=m^2_V,
\ena
where $N_c=3$ is the number of colors. As before we have 
$w_i = m_{q_i}/(m_{q_1}+m_{q_2})$. Further $O_{P}^{\mu}=\gamma^{\mu}$ and
$O_{V}^{\mu}=\gamma^{\mu}\gamma_{5}$.

The compositeness conditions (\ref{compcon}) and the fit to the
leptonic decay constants (\ref{eq:lept}) provide $2n_{H}$ constraint
equations for $2n_{H}+5$ model parameters. As further constraints on the 
parameter space, we have decided to fit the model to the 
eight fundamental electromagnetic decays listed in Table~\ref{tab:em-widths}.
We refer the reader to Ref.\,\cite{Branz:2009cd} for the details of the
one--loop calculation of the electromagnetic decays. The results of
the (overconstrained) least--squares fit to the leptonic decay constants 
and the electromagnetic decay widths and the corresponding 
experimental/theoretical input values can be found in 
Tables~\ref{tab:leptonic} and Table~\ref{tab:em-widths}, respectively. The
agreement between the fit values and input values is quite satisfactory. 
   
\begin{table}[t]
\caption{Input values for the leptonic decay constants $f_H$ (in MeV) and
our least-squares fit values.}
\label{tab:leptonic}
\begin{center}
\def\arraystretch{1.5}
\begin{tabular}{cccc|}
\hline\hline
    & Fit Values  & Other &  Ref.  \\
\hline
$f_\pi$  & 128.7 & $130.4\,\pm\, 0.2 $   & \cite{PDG,RosnerStone}\\
%
%
$f_K$   & 156.1 & $156.1 \,\pm\, 0.8 $  & \cite{PDG,RosnerStone}\\
%
%
$f_{D}$  & 205.9 & $206.7 \,\pm\, 8.9 $ & \cite{PDG,RosnerStone}\\
%
%
$f_{D_s}$ & 257.5 & $257.5 \,\pm\, 6.1 $ & \cite{PDG,RosnerStone}\\
%
%
$f_{B}$ & 191.1 & $192.8 \,\pm\, 9.9  $ & \cite{LatticePRD81}\\
%
%
$f_{B_s}$ & 234.9 & $238.8 \,\pm\, 9.5 $ & \cite{LatticePRD81}\\
%
%
$f_{B_c}$ & 489.0 & $489 \,\pm\, 5 $ & \cite{LatticeTWQCD}\\
%
%
$f_{\rho}$ & 221.1 & $221 \,\pm\, 1 $ & \cite{PDG}\\
\hline\hline
\end{tabular}
\begin{tabular}{|cccc}
\hline\hline
    & This work  &  Other &  Ref.  \\
\hline
$f_\omega$  & 198.5  & $198 \,\pm\, 2 $ & \cite{PDG} \\
%
%
$f_\phi$    & 228.2  & $227\,\pm\,  2 $ & \cite{PDG} \\
%
%
$f_{\Jpsi}$  & 415.0  & $415\,\pm\, 7  $ & \cite{PDG} \\
%
%
$f_{K^\ast}$  & 213.7  & $217\,\pm\, 7  $ & \cite{PDG} \\
%
%
$f_{D^\ast}$   & 243.3  & $245\,\pm\, 20  $ & \cite{Lubicz} \\
%
%
$f_{D^\ast_s}$   & 272.0  & $272\,\pm\,26  $ & \cite{Lubicz} \\
%
%
$f_{B^\ast}$   & 196.0  & $196\,\pm\, 44 $ & \cite{Lubicz} \\
%
$f_{B_s^\ast}$   & 229.0  & $229\,\pm\, 46  $ & \cite{Lubicz} \\
\hline\hline
\end{tabular}
\end{center}
\end{table}

\begin{table}[ht]
\begin{center}
\def\arraystretch{1.5}
\caption{Input values for some basic electromagnetic decay widths and our 
least-squares 
fit values (in keV).}
\label{tab:em-widths}
\vspace*{0.2cm}
\begin{tabular}{lcc}
\hline\hline 
Process & Fit Values & Data~\cite{PDG}  \\
\hline
$\pi^0\to\gamma\gamma$          & \,\,  $ 5.06 \times 10^{-3}$ \,\,& 
                                \,\,$(7.7 \pm 0.4) \times 10^{-3}$\,\,\\ 
$\eta_c\to\gamma\gamma$         & 1.61 & 1.8 $\pm$ 0.8 \\ 
$\rho^{\pm}\to\pi^{\pm}\gamma$    & 76.0     &  67 $\pm$ 7  \\
$\omega\to\pi^0\gamma$          & 672      &  703 $\pm$ 25    \\
$K^{\ast \pm}\to K^\pm\gamma$      & 55.1     &  50 $\pm$ 5          \\
$K^{\ast 0}\to K^0\gamma$         & 116      &  116 $\pm$ 10      \\
$D^{\ast \pm}\to D^\pm\gamma$      & 1.22     &  1.5 $\pm$ 0.5 \\
$\Jpsi \to \eta_c \gamma $      & 1.43     &  1.58 $\pm$ 0.37  \\
\hline
\end{tabular}
\end{center}
\end{table}
The results of the fit for the values of quark masses $m_{q_{i}}$, the 
infrared cutoff parameter $\lambda$
and the size parameters $\Lambda_{H_{i}}$are given in Eqs.~(\ref{eq: fitmas}),
(\ref{eq:fitsize1}) and (\ref{eq:fitsize2}), respectively.

\be
\def\arraystretch{2}
\begin{array}{cccccc}
     m_u        &      m_s        &      m_c       &     m_b & \lambda  &   
\\\hline
 \ \ 0.235\ \   &  \ \ 0.424\ \   &  \ \ 2.16\ \   &  \ \ 5.09\ \   & 
\ \ 0.181\ \   & \ {\rm GeV} 
\end{array}
\label{eq: fitmas}
\en

\be
\def\arraystretch{2}
\begin{array}{ccccccccc}
 \Lambda_\pi   & \Lambda_K   & \Lambda_D        &  \Lambda_{D_s} &
 \Lambda_{B} & \Lambda_{B_s} & \Lambda_{B_c} &  \Lambda_{\rho}    & 
\\\hline
\ \ 0.87 \ \  & \ \ 1.04 \ \ & \ \ 1.47\ \  & \ \ 1.57 \ \ &
\ \ 1.88 \ \  & \ \ 1.95 \ \ & \ \ 2.42\ \  & \ \ 0.61 \ \ & \  {\rm GeV}
\end{array}
\label{eq:fitsize1}
\en

\be
\def\arraystretch{2}
\begin{array}{ccccccccc}
 \Lambda_\omega   & \Lambda_\phi   & \Lambda_{\Jpsi} &  \Lambda_{K^\ast} &
 \Lambda_{D^\ast}  & \Lambda_{D^\ast_s} & \Lambda_{B^\ast} &  \Lambda_{B^\ast_s}    & 
\\\hline
\ \ 0.47 \ \  & \ \ 0.88 \ \ & \ \ 1.48\ \ & \ \ 0.72 \ \ &
\ \ 1.16 \ \  & \ \ 1.17 \ \ & \ \ 1.72\ \ & \ \ 1.71 \ \ & \  {\rm GeV}
\end{array}
\label{eq:fitsize2}
\en
The constituent
quark masses and the values of the size parameter
fall into the expected range. The size parameters show the expected general 
pattern in that the geometrical sizes of the mesons 
$\propto \Lambda_{H_{i}}^{-1}$ shrink as their masses increase.

The present numerical least-squares fit and the values for the model 
parameters supersede the results of a similar analysis
given in \cite{Branz:2009cd} which used a different set of
electromegnetic decays in the fit. In the present fit we have also updated
some of the theoretical/experimental input values.

\section{Transition Form Factors}
\label{sec:tff}
Given the fact that all model parameters have been fixed, the covariant 
quark model can now be utilized to calculate any given decay process in a 
parameter-free way. 
As a first application we calculate the form factors
describing the transitions of heavy $B(B_s)$ mesons into
light mesons, e.g. $B, B_{s}\to \pi,K,\rho,K^\ast,\phi.$ These quantities
are of great interest due to their applications in semileptonic,
nonleptonic and rare decays of the $B$ and $B_s$ mesons. 
They have been calculated within the LCSR approach
in the region of large recoil (small momentum transfer) and have been
extrapolated to the low recoil region.
Our approach allows one to evaluate the form factors in the full
kinematical range including the near-zero recoil region.

Below, we list the definitions of the dimensionless invariant transition
form factors together with the covariant quark model expressions that allow 
one to calculate them. We closely follow the notation used in our 
papers \cite{IKS}.

\bea
&&
\langle 
P^{\,\prime}_{[\bar q_3 q_2]}(p_2)\,|\,\bar q_2\, O^{\,\mu}\, q_1\,| P_{[\bar q_3 q_1]}(p_1)
\rangle
\nn
&=&
N_c\, g_P\,g_{P^{\,'}}\!\!  \int\!\! \frac{d^4k}{ (2\pi)^4 i}\, 
\widetilde\Phi_P\Big(-(k+w_{13})^2\Big)\,
\widetilde\Phi_{P^{\,'}}\Big(-(k+w_{23})^2\Big)
{\rm tr} \biggl[
O^{\,\mu}\, S_1(k+p_1)\, \gamma^5\, S_3(k)\, \gamma^5\, S_2(k+p_2) 
\biggr]
\nn
 & = & F_+(q^2)\, P^{\,\mu} + F_-(q^2)\, q^{\,\mu}\,,
\label{eq:PP'}
\ena

\bea
&&
\langle 
P^{\,\prime}_{[\bar q_3 q_2]}(p_2)\,
|\,\bar q_2\, (\sigma^{\,\mu\nu}q_\nu) \, q_1\,| 
P_{[\bar q_3 q_1]}(p_1)
\rangle
\,=\,
\nn
&=&
N_c\, g_P\,g_{P^{\,'}} \!\! \int\!\! \frac{d^4k}{ (2\pi)^4 i}\, 
\widetilde\Phi_P\Big(-(k+w_{13})^2\Big)\,
\widetilde\Phi_{P^{\,'}}\Big(-(k+w_{23})^2\Big)
{\rm tr} \biggl[ 
\sigma^{\,\mu\nu}q_\nu \,
S_1(k+p_1)\, \gamma^5\, S_3(k)\, \gamma^5 \,S_2(k+p_2) 
\biggr]
\nn
 & = &
\frac{i}{m_1+m_2}\,\left(q^2\, P^{\,\mu}-q\cdot P\, q^{\,\mu}\right)\,F_T(q^2),
\label{eq:PP'T}
\ena

\bea
&&
\langle 
V(p_2,\epsilon_2)_{[\bar q_3 q_2]}\,
|\,\bar q_2\, O^{\,\mu}\,q_1\, |\,P_{[\bar q_3 q_1]}(p_1)
\rangle 
\nn
&=&
N_c\, g_P\,g_V \!\! \int\!\! \frac{d^4k}{ (2\pi)^4 i}\, 
\widetilde\Phi_P\Big(-(k+w_{13})^2\Big)\,
\widetilde\Phi_V\Big(-(k+w_{23})^2\Big)
{\rm tr} \biggl[ 
O^{\,\mu} \,S_1(k+p_1)\,\gamma^5\, S_3(k) \not\!\epsilon_2^{\,\,\dagger} \,
S_2(k+p_2)\, \biggr]
\nn
 & = &
\frac{\epsilon^{\,\dagger}_{\,\nu}}{m_1+m_2}\,
\left( - g^{\mu\nu}\,P\cdot q\,A(q^2) + P^{\,\mu}\,P^{\,\nu}\,A_+(q^2)
       + q^{\,\mu}\,P^{\,\nu}\,A_-(q^2) 
+ i\,\varepsilon^{\mu\nu\alpha\beta}\,P_\alpha\,q_\beta\,V(q^2)\right),
\label{eq:PV}
\ena

\bea
&&
\langle 
V(p_2,\epsilon_2)_{[\bar q_3 q_2]}\,
|\,\bar q_2\, (\sigma^{\,\mu\nu}q_\nu(1+\gamma^5))\,q_1\, |\,P_{[\bar q_3 q_1]}(p_1)
\rangle 
\nn
&=&
N_c\, g_P\,g_V \!\! \int\!\! \frac{d^4k}{ (2\pi)^4 i}\, 
\widetilde\Phi_P\Big(-(k+w_{13})^2\Big)\,
\widetilde\Phi_V\Big(-(k+w_{23})^2\Big)
{\rm tr} \biggl[ 
(\sigma^{\,\mu\nu}q_\nu(1+\gamma^5))
\,S_1(k+p_1)\,\gamma^5\, S_3(k) \not\!\epsilon_2^{\,\,\dagger} \,S_2(k+p_2)\, 
\biggr]
\nn
 & = &
\epsilon^{\,\dagger}_{\,\nu}\,
\left( - (g^{\mu\nu}-q^{\,\mu}q^{\,\nu}/q^2)\,P\cdot q\,a_0(q^2) 
       + (P^{\,\mu}\,P^{\,\nu}-q^{\,\mu}\,P^{\,\nu}\,P\cdot q/q^2)\,a_+(q^2)
+ i\,\varepsilon^{\mu\nu\alpha\beta}\,P_\alpha\,q_\beta\,g(q^2)\right).
\label{eq:PVT}
\ena
We use $P=p_1+p_2$ and $q=p_1-p_2$ and the on--shell conditions 
$\epsilon_2^\dagger\cdot p_2=0$,
$p_i^2=m_i^2$. Since there are three quark species involved in the transition,
we have introduced a two--subscript notation
$w_{ij}=m_{q_j}/(m_{q_i}+m_{q_j})$ $(i,j=1,2,3)$ such that $w_{ij}+w_{ji}=1$. 
The form factors defined in Eq.\,(\ref{eq:PVT}) satisfy the physical 
requirement $a_0(0)=a_+(0)$, which ensures that no kinematic singularity 
appears in the matrix element at $q^2=0$.  
For reference, it is useful to relate the above form factors  
to those used in Ref.\,\cite{Khodjamirian:2006st}. 
The relations read 
\bea
\label{cformfactors}
F_+ &=& f_+\,,\quad 
F_- = -\,\frac{m_1^2-m_2^2}{q^2}\,(f_+ - f_0)\,, \quad 
F_T = f_T\,, 
\nn
&&\nn
A_0 &=& \frac{m_1 + m_2}{m_1 - m_2}\,A_1\,, \quad 
A_+ = A_2\,,\quad
A_- =  \frac{2m_2(m_1+m_2)}{q^2}\,(A_3 - A_0)\,, \quad
V = V\,, 
\nn
&&\nn
a_0 &=& T_2\,, \quad g = T_1\,, \quad
a_+  =  T_2+\frac{q^2}{m_1^2-m_2^2}\,T_3\,.
\ena
The form factors (\ref{cformfactors}) satisfy
the constraints
\bea
A_0(0)&=&A_3(0)\\ \nonumber
2m_2A_3(q^2) &=& (m_1+m_2) A_1(q^2) -(m_1-m_2) A_2(q^2)\,.
\ena

\begin{figure}[ht]
\begin{center}
\hspace*{-0.5cm}
\begin{tabular}{lr}
\includegraphics[width=0.40\textwidth]{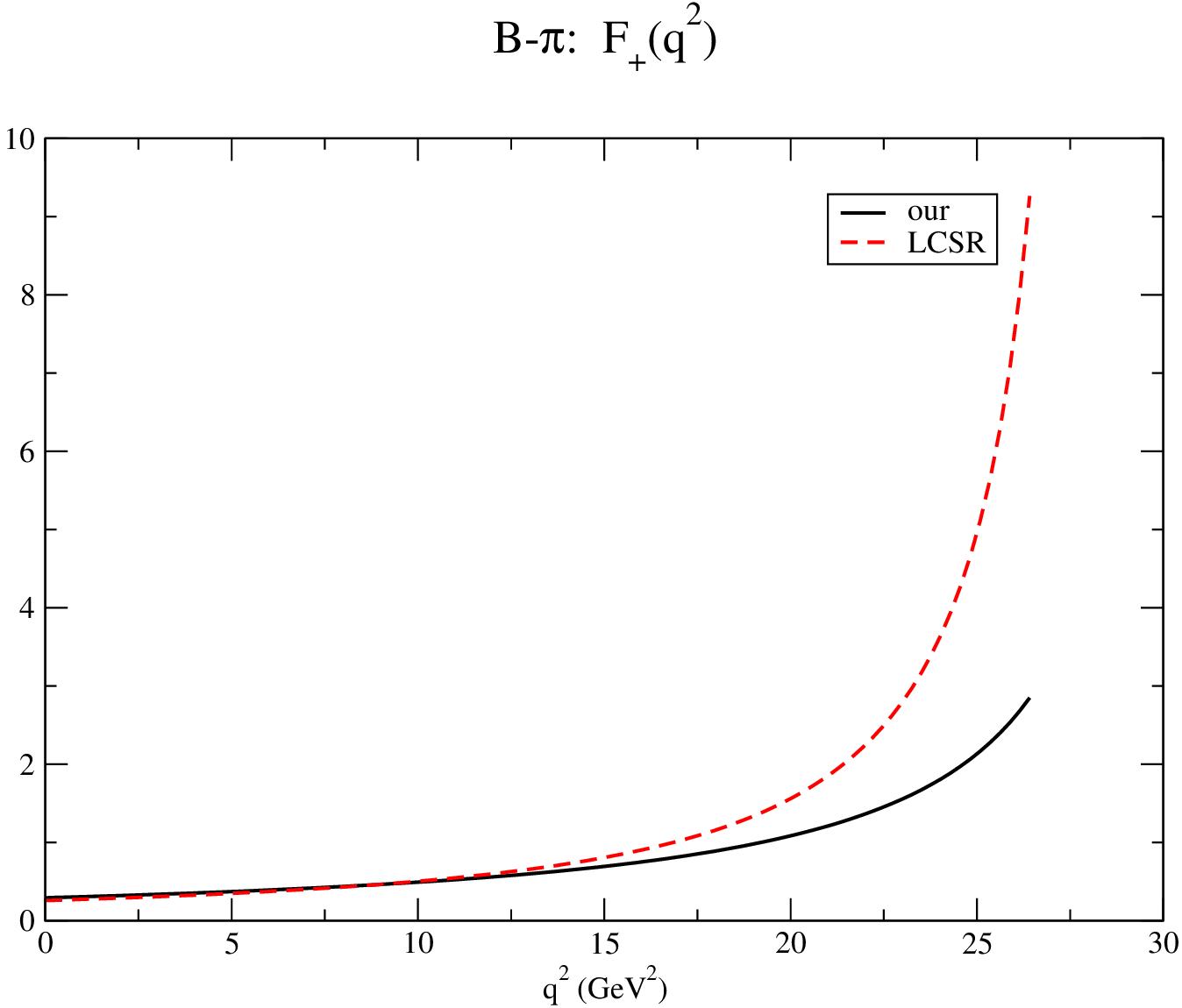} & \hspace{0.6truecm}
\includegraphics[width=0.40\textwidth]{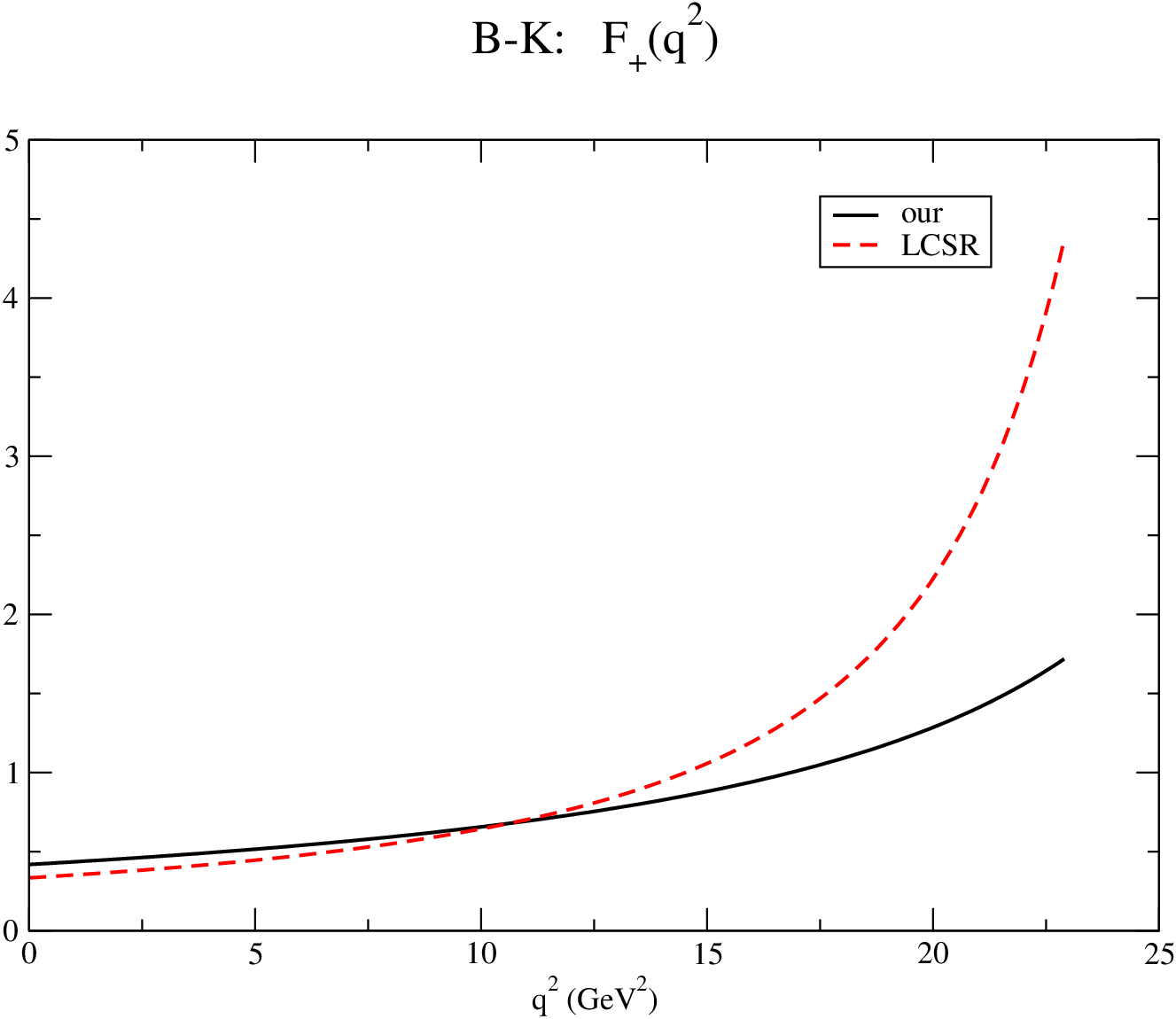} \\[2ex]
\includegraphics[width=0.40\textwidth]{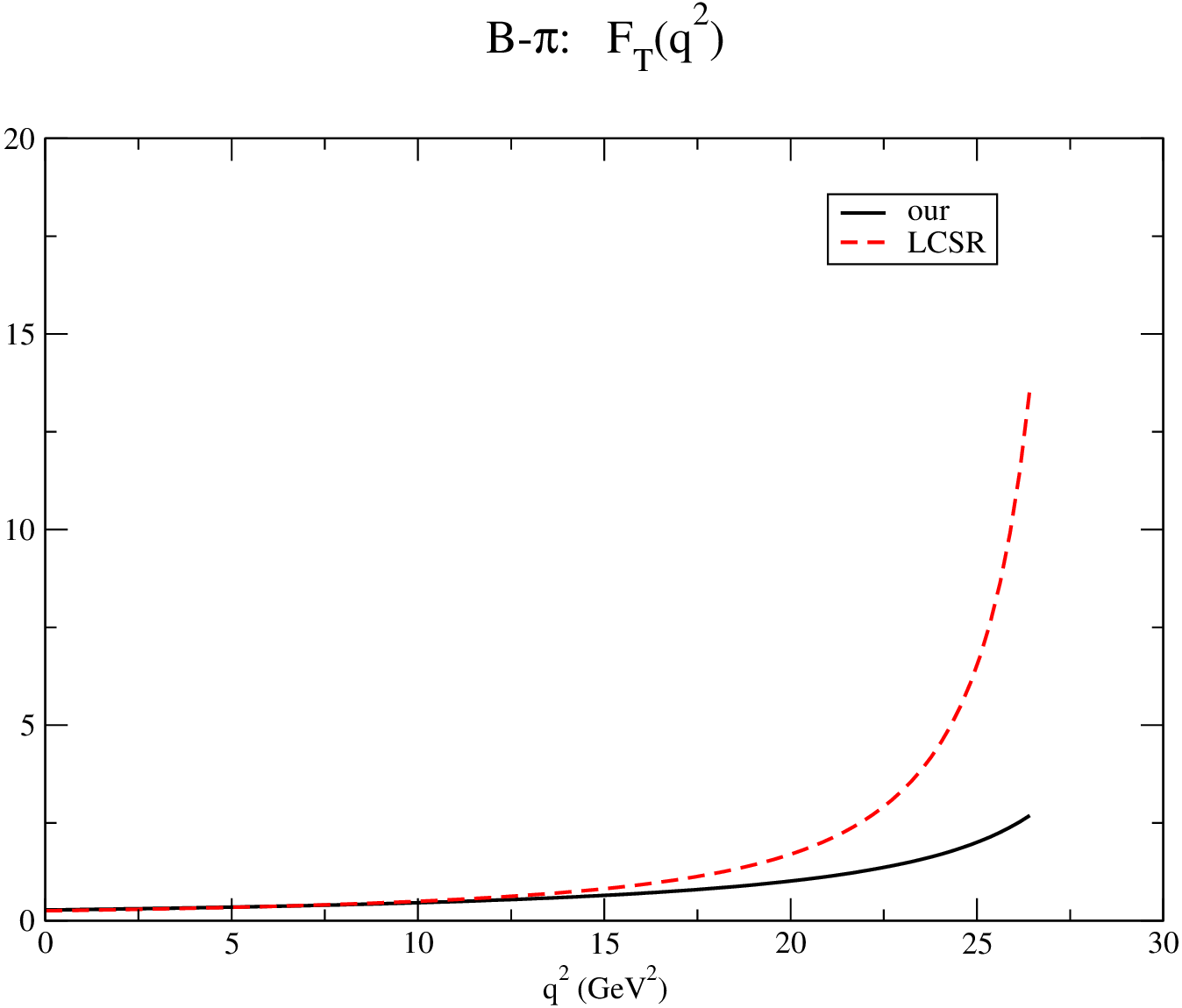} &
\includegraphics[width=0.40\textwidth]{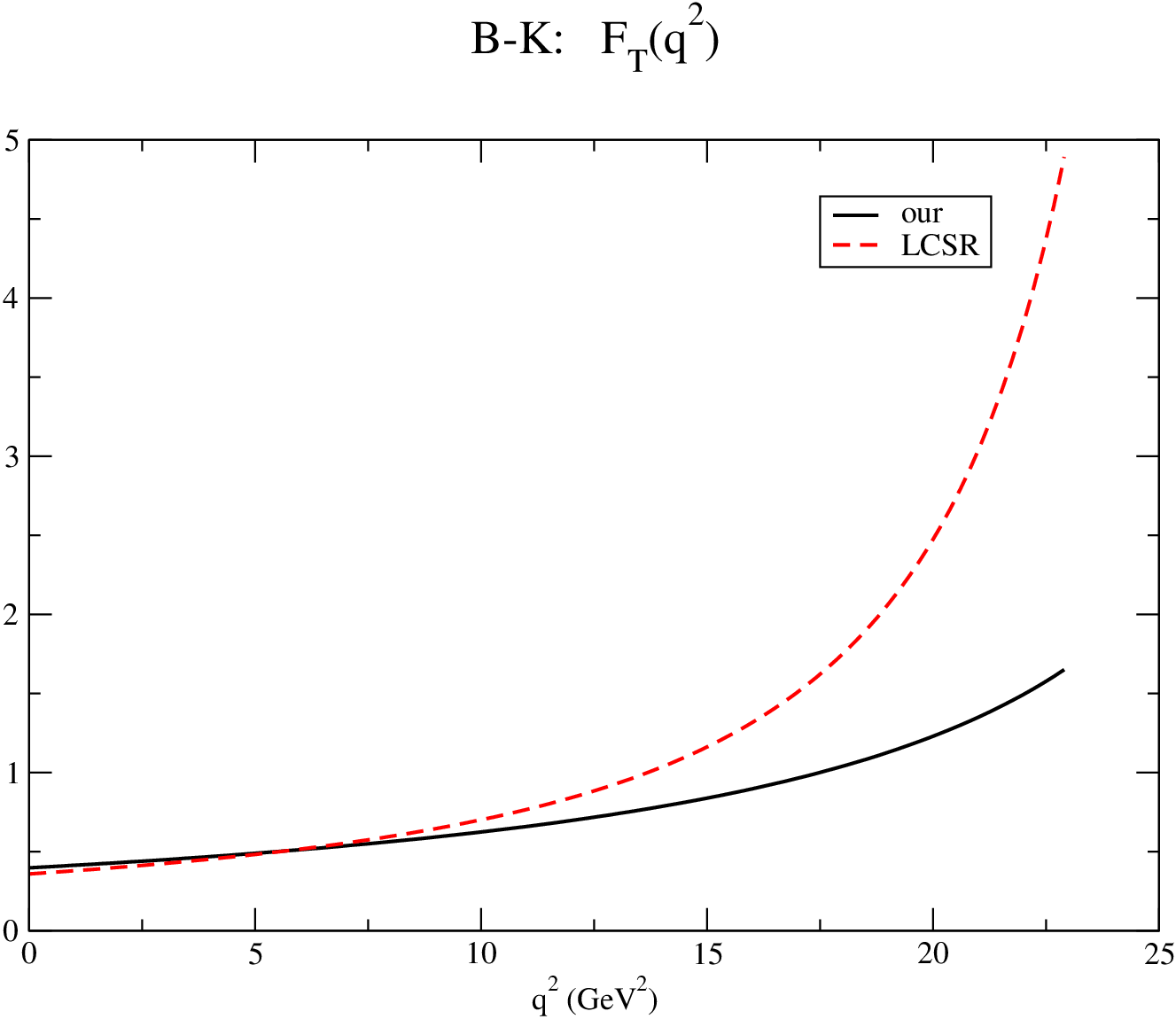}
\end{tabular}
\end{center}
\caption{\label{fig:ff-BP}
Our results for the form factors appearing 
in Eqs.\,(\protect\ref{eq:PP'}) and (\protect\ref{eq:PP'T}) -- \emph{Left panel},  
$B-\pi-$transition; 
                                and \emph{right panel}; 
$B-K$transition. 
For comparison, we plot the corresponding LCSR curves
from Ref.~\protect\cite{Ball}.
}
\end{figure}

\begin{figure}[ht]
\begin{center}
\hspace*{-0.5cm}
\begin{tabular}{lr}
\includegraphics[width=0.40\textwidth]{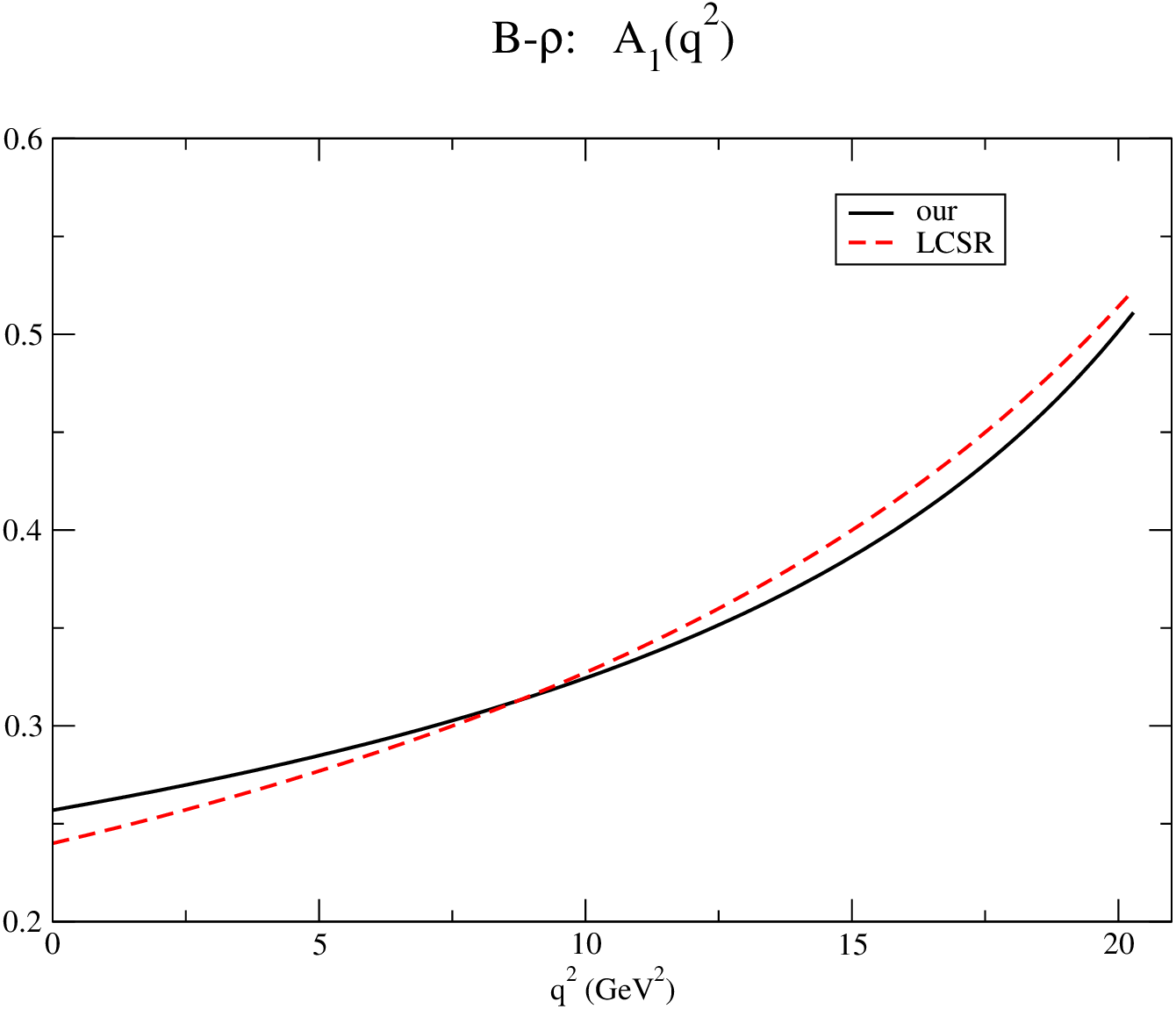} & \hspace{0.6truecm}
\includegraphics[width=0.40\textwidth]{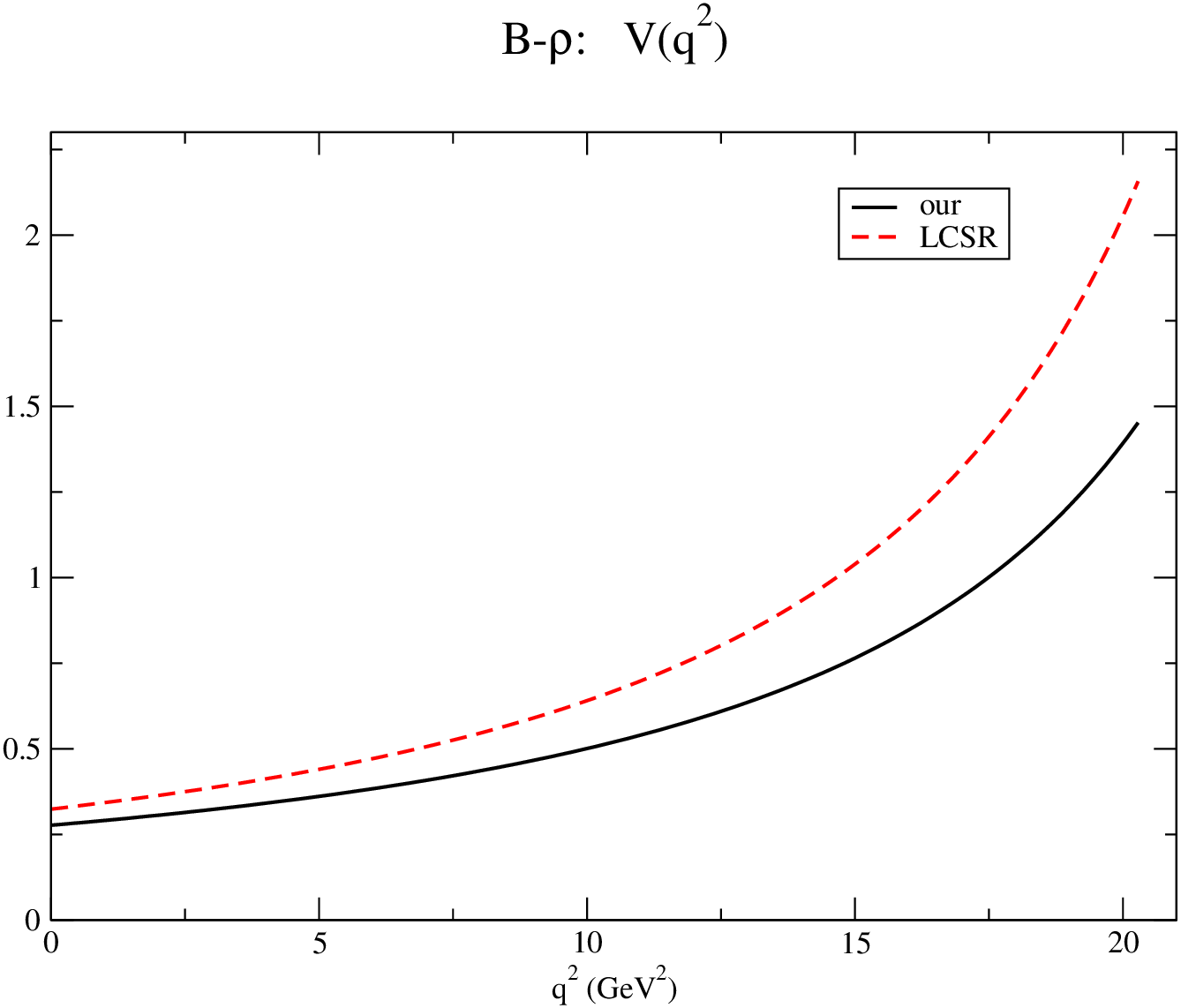} \\[2ex]
\includegraphics[width=0.40\textwidth]{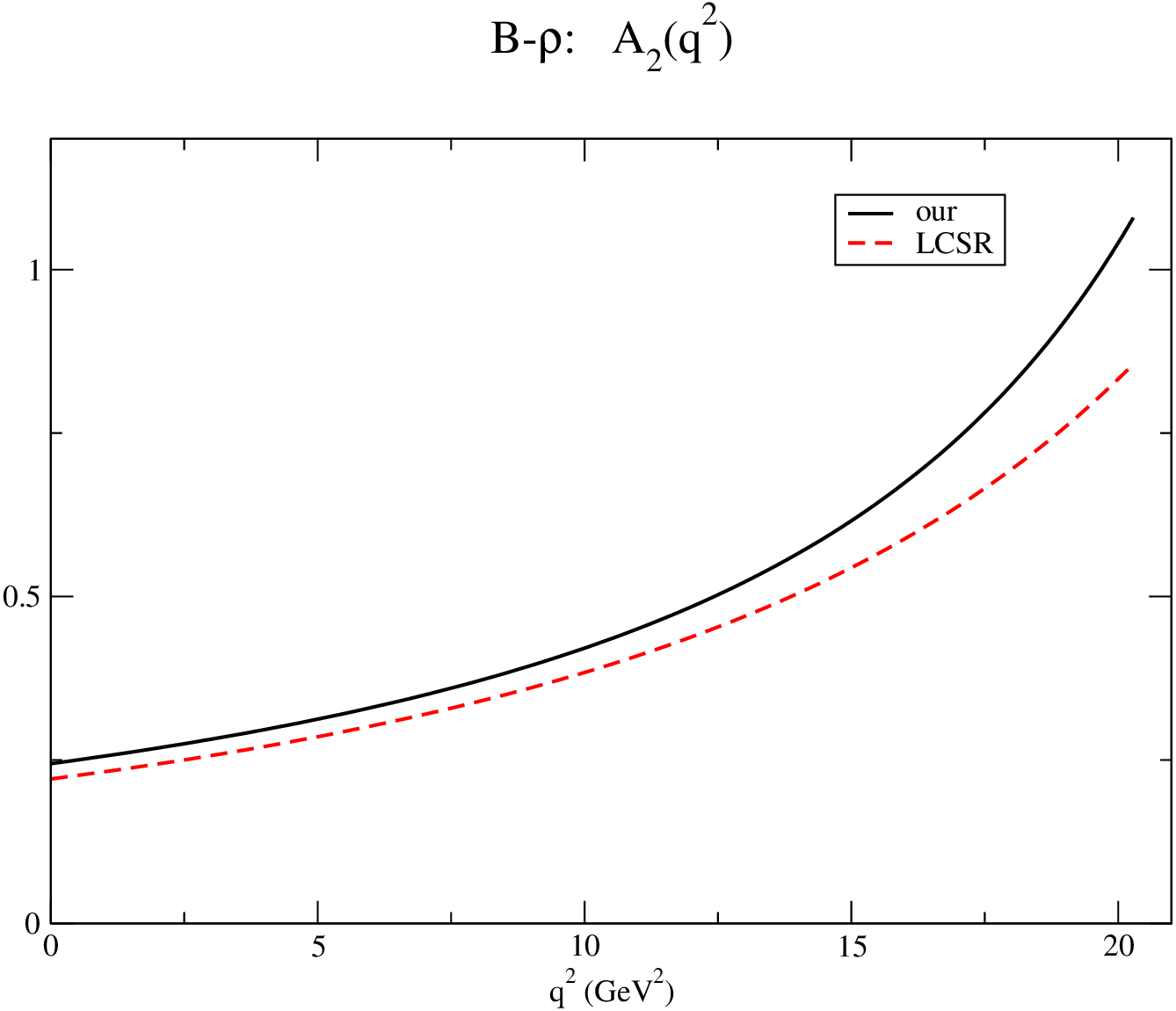} &
\includegraphics[width=0.40\textwidth]{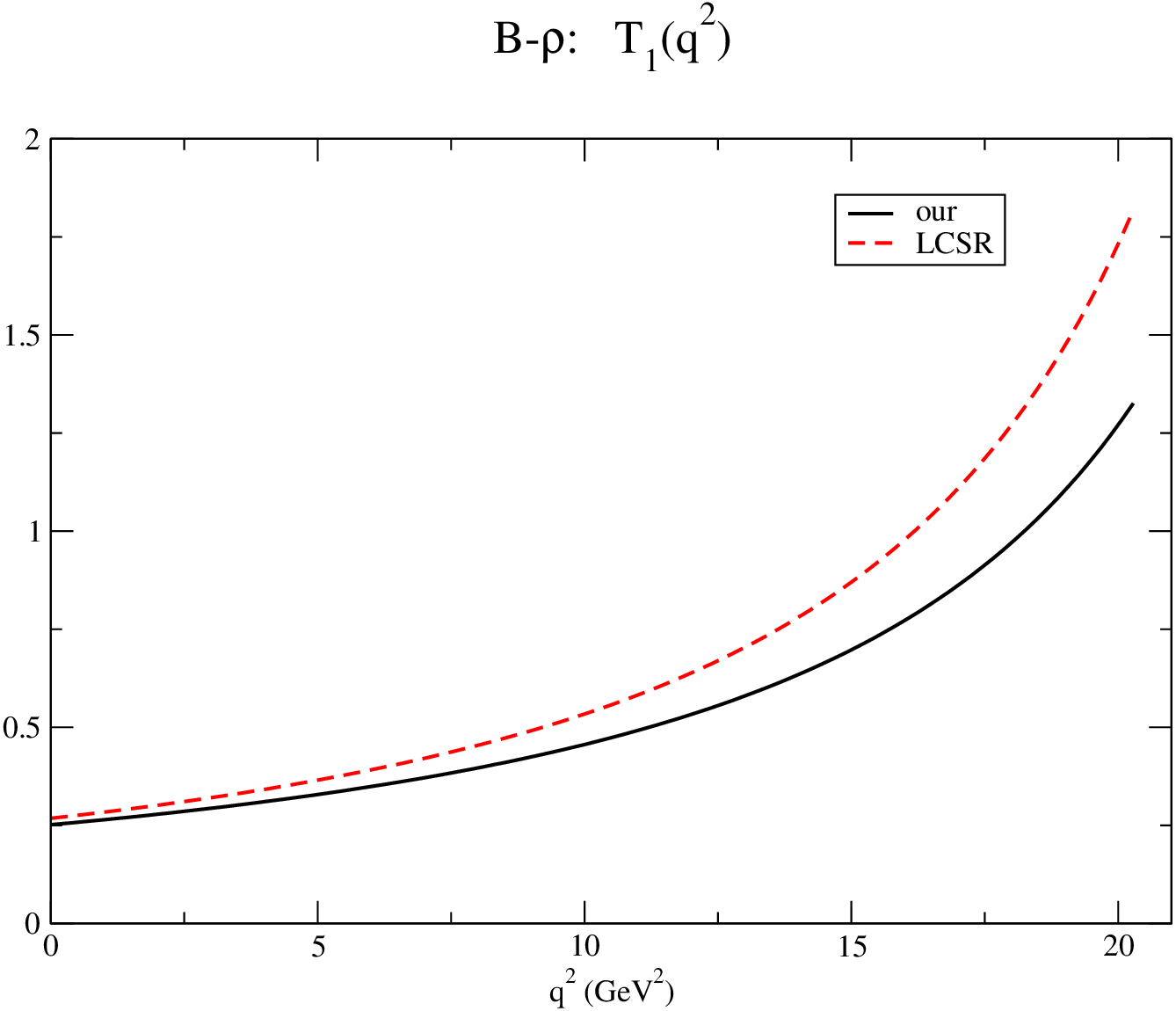}
\end{tabular}
\end{center}
\caption{\label{fig:ff-Brho}
Our results for the form factors appearing 
in Eqs.\,(\protect\ref{eq:PV}) and (\protect\ref{eq:PVT}) for  $B-\rho$ transition. 
For comparison, we plot the corresponding LCSR curves
from Ref.~\protect\cite{Ball}. }
\end{figure}

\begin{figure}[ht]
\begin{center}
\hspace*{-0.5cm}
\begin{tabular}{lr}
\includegraphics[width=0.40\textwidth]{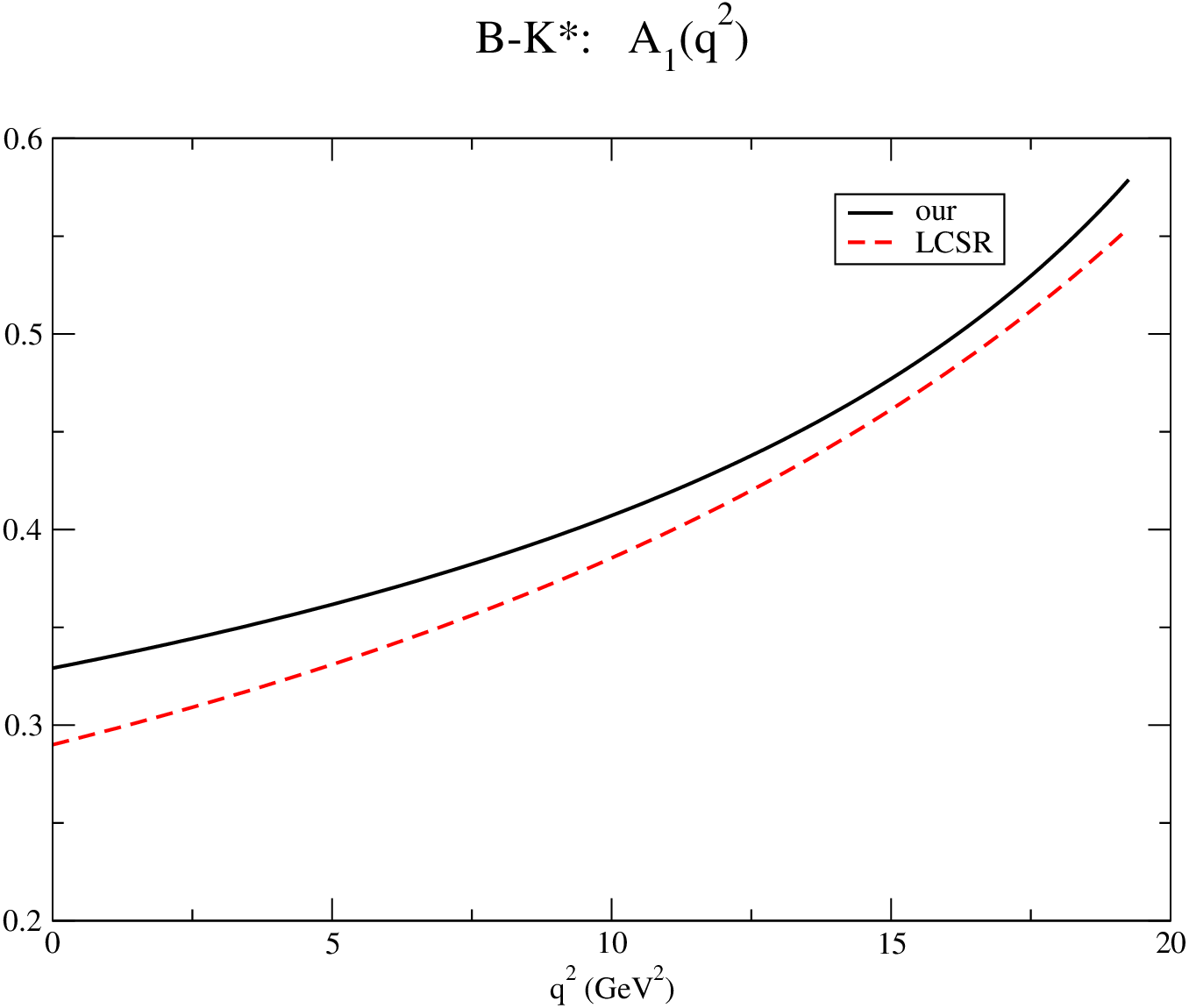} & \hspace{0.6truecm}
\includegraphics[width=0.40\textwidth]{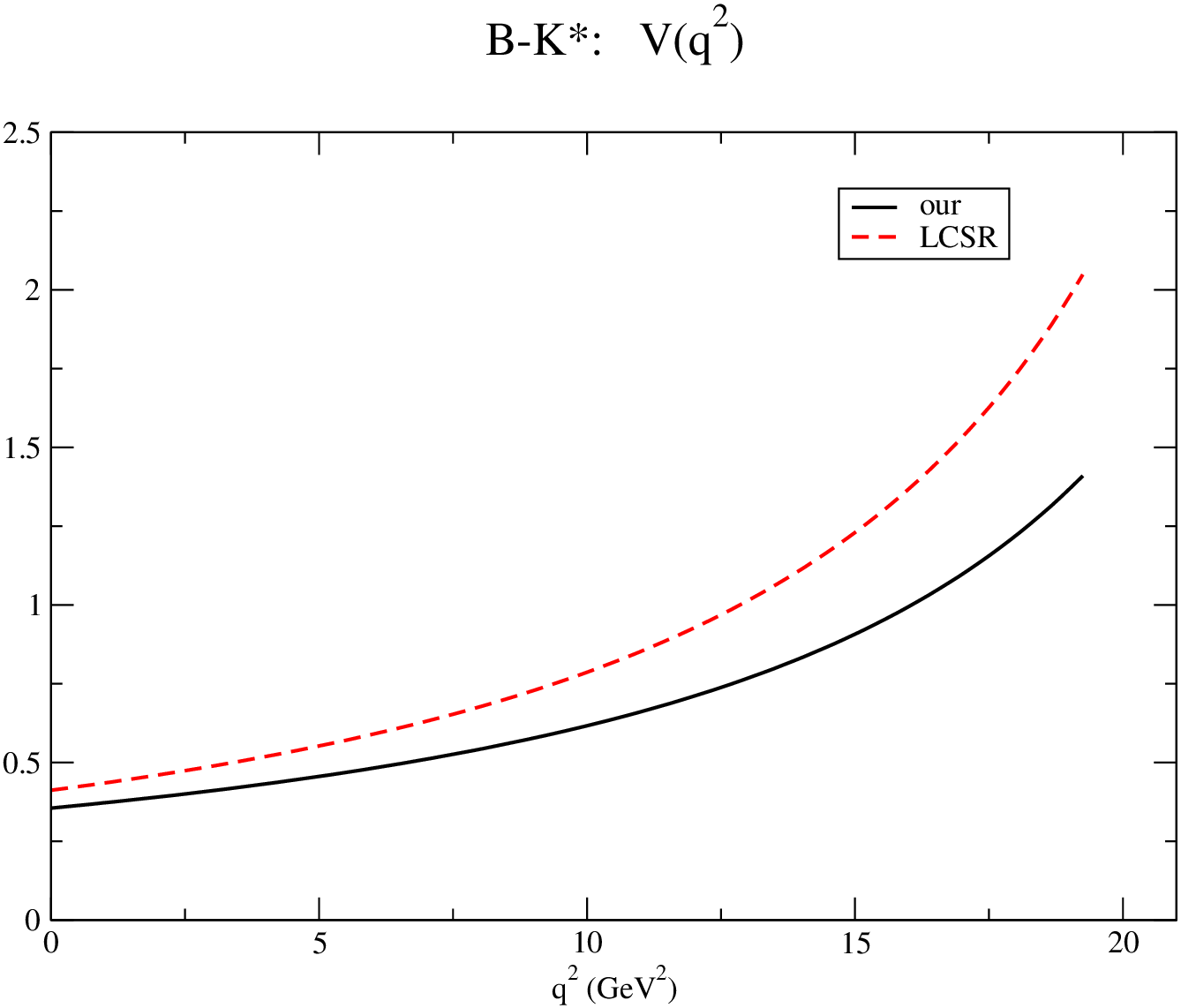} \\   [2ex]
\includegraphics[width=0.40\textwidth]{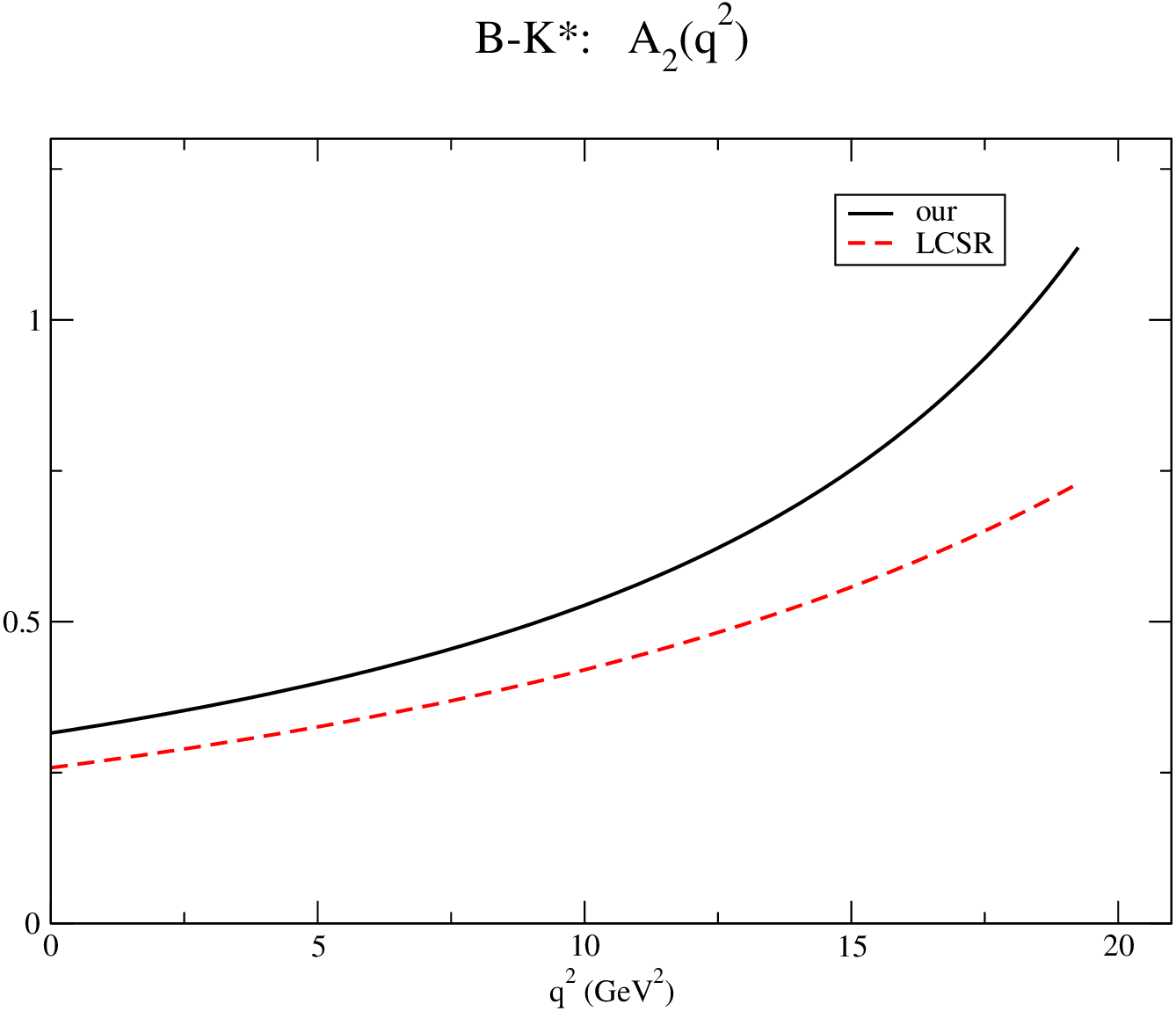} &
\includegraphics[width=0.40\textwidth]{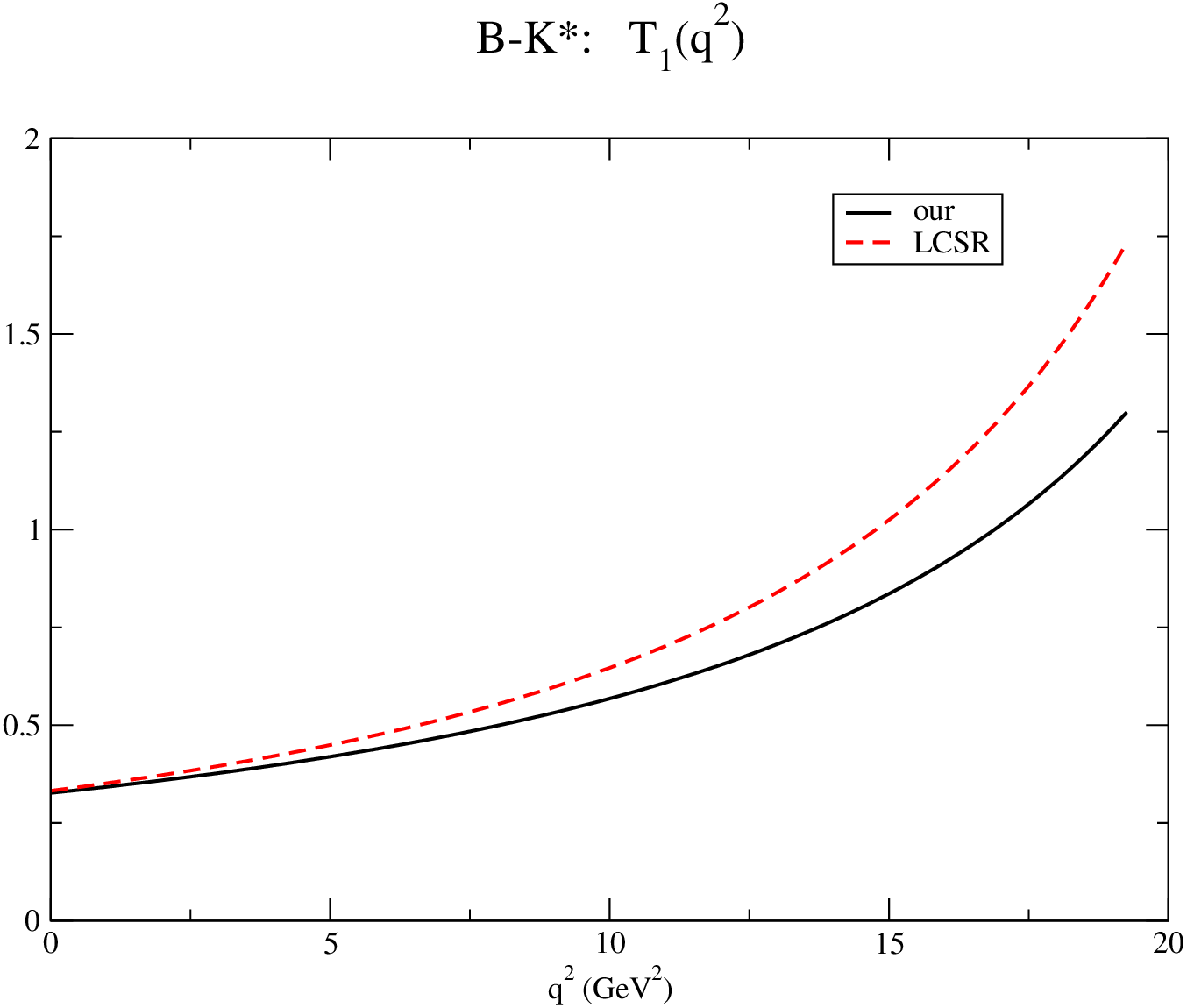}
\end{tabular}
\end{center}
\caption{\label{fig:ff-BKv}
Our results for the form factors appearing 
in Eqs.\,(\protect\ref{eq:PV})  and (\protect\ref{eq:PVT}) for  $B-K^\ast$ transition. 
For comparison, we plot the corresponding LCSR curves 
from Ref.~\protect\cite{Ball}. }
\end{figure}

\begin{figure}[ht]
\begin{center}
\hspace*{-0.5cm}
\begin{tabular}{lr}
\includegraphics[width=0.40\textwidth]{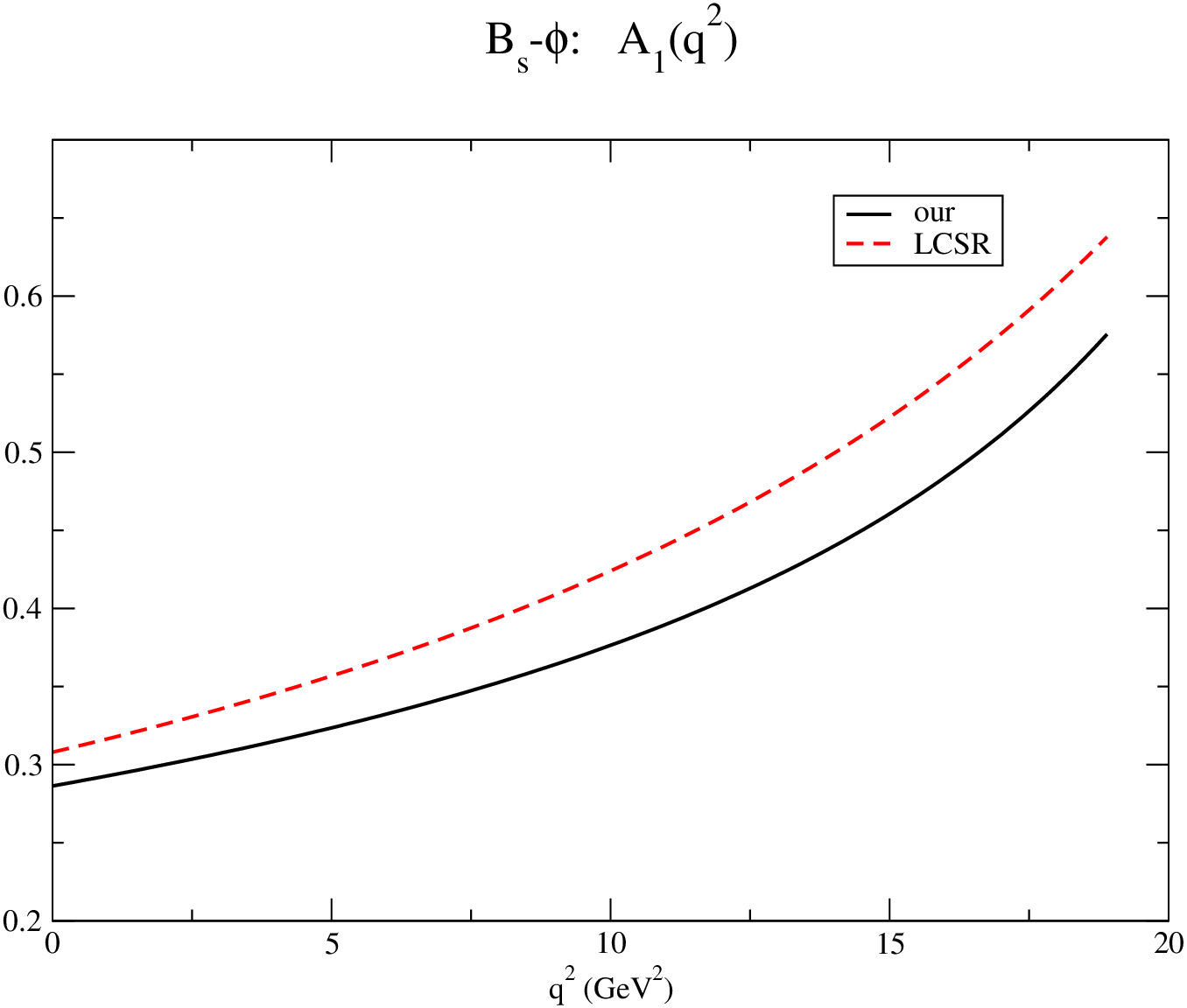} & \hspace{0.6truecm}
\includegraphics[width=0.40\textwidth]{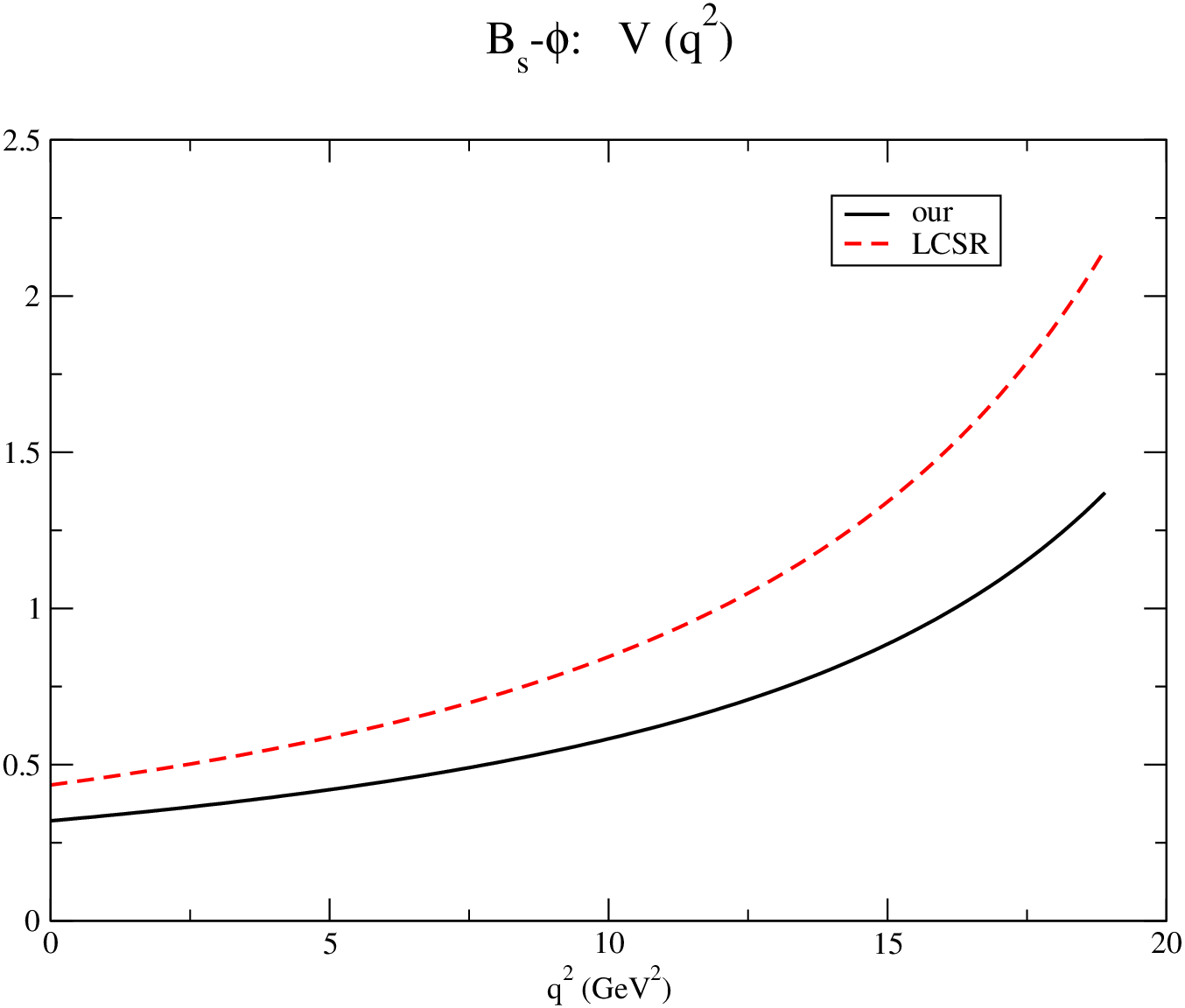} \\[2ex]
\includegraphics[width=0.40\textwidth]{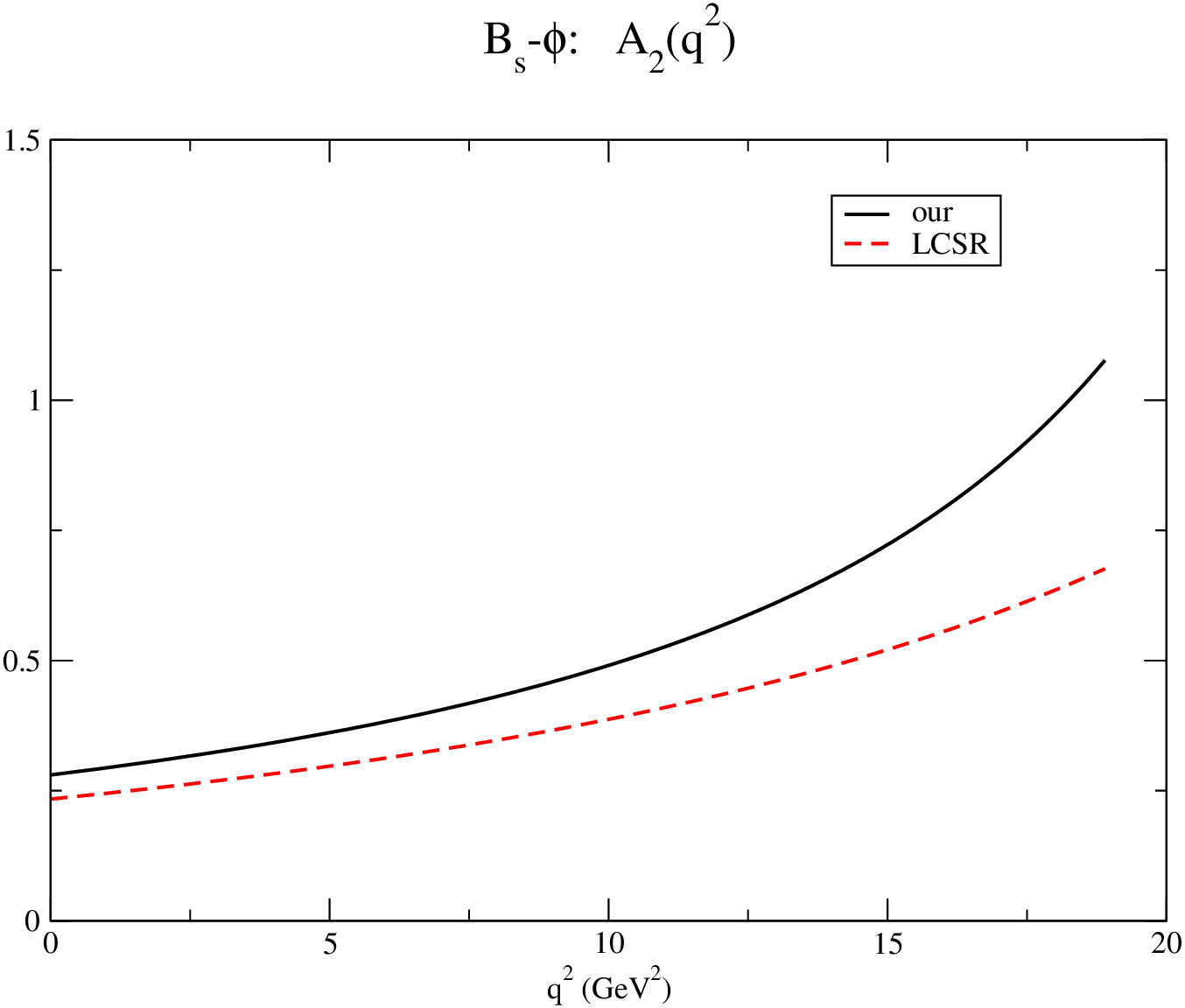} &
\includegraphics[width=0.40\textwidth]{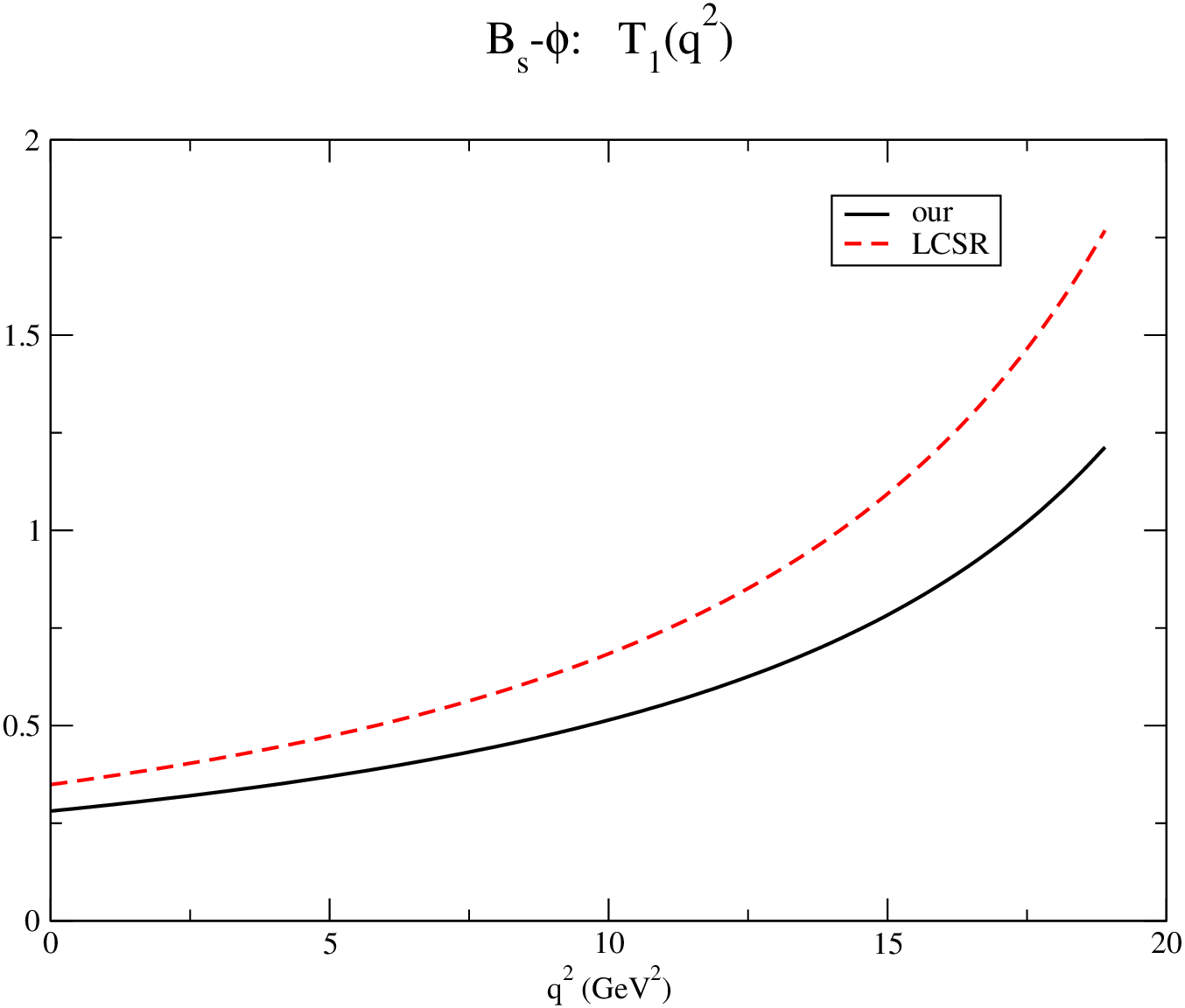}
\end{tabular}
\end{center}
\caption{\label{fig:ff-BsPhi}
Our results for the form factors appearing 
in Eqs.\,(\protect\ref{eq:PV}) 
      and (\protect\ref{eq:PVT}) for  $B_s-\phi$ transition. 
For comparison, we plot the corresponding LCSR curves
from Ref.~\protect\cite{Ball}. }
\end{figure}

In Figs.~\ref{fig:ff-BP}-\ref{fig:ff-BsPhi},
we plot our calculated form factors in the entire kinematical range
$0\le q^2\le q^2_{\rm max}$. For comparison, we also show the results
obtained from the light-cone sum rules analysis \cite{Ball}.
In Table~\ref{tab:semileptonic}, we collect our predictions for
the form factors at the maximum recoil point $q^{2}=0$ and provide
a comparison with results obtained within other approaches.
The figures and tables highlight the wide range of phenomena accessible
within our approach.

\begin{table}[ht]
\caption{ $q^2=0$ results for the transition form factors in various model
approaches. }
\label{tab:semileptonic}
\begin{center}
\def\arraystretch{1.5}
\begin{tabular}{lcccccccc}
\hline\hline
  &\,\, This work\,\, & \,LCSR-1 \cite{Ball}\, & 
 \,LCSR-2  \cite{Khodjamirian:2006st}\, &
 \,\, DSE  \cite{Ivanov:2007cw}\,\, & QCD SR \cite{Colangelo:1995jv} & 
\,\, DQM \cite{Melikhov}\,\, & \,\,RQM \cite{Ebert:2006nz}\,\, &
\,\,RCQM \cite{Faessler:2002ut}\,\,\\
\hline 
 $F_+^{B\pi}(0)$   & 0.29 & 0.258$\pm$0.031\,\,  &\,\, 0.25$\pm$0.05\,\,  & 
\,\,0.24$\pm$0.05\,\, & \,\,0.24$\pm$0.03\,\, & 0.29 & 0.22 & 0.27\\ 
 $F_+^{B K}(0)$    &  0.42  & 0.335$\pm$0.042  & 0.31$\pm$0.04 & 0.30$\pm$0.06 &
 0.25$\pm$0.03 & 0.36 &  & 0.36 \\ 
 $F_T^{B\pi}(0)$   & 0.27  & 0.253$\pm$0.028 & 0.21$\pm$0.04 & 0.25$\pm$0.05 & &
 0.28 & & \\ 
 $F_T^{B K}(0)$    & 0.40 & 0.359$\pm$0.038  & 0.27$\pm$0.04 & 0.32$\pm$0.06 & 
0.14$\pm$0.03 & 0.35 & & 0.34\\  
 $V^{B \rho}(0)$   & 0.28 &  0.324$\pm$0.029  & 0.32$\pm$0.10 &  0.31$\pm$0.06 &&
 0.31 & 0.30 &\\
 $V^{B K^*}(0)$    & 0.36 & 0.412$\pm$0.045 & 0.39$\pm$0.11 &  0.37$\pm$0.07 & 
0.47$\pm$0.03 & 0.44 & &\\ 
 $V^{B_s \phi}(0)$    & 0.32 & 0.434$\pm$0.035 & & & & & &\\ 
 $A_1^{B \rho}(0)$ & 0.26  & 0.240$\pm$0.024 & 0.24$\pm$0.08 & 0.24$\pm$0.05 & &
 0.26 & 0.27 &\\
 $A_1^{B K^*}(0)$  & 0.33 & 0.290$\pm$0.036 & 0.30$\pm$0.08 &  0.29$\pm$0.06 & 
0.37$\pm$0.03 & 0.36 & &\\
 $A_1^{B_s\phi}(0)$  & 0.29 & 0.311$\pm$0.029  & & & & & &\\
 $A_2^{B \rho}(0)$ & 0.24 & 0.221$\pm$0.023 & 0.21$\pm$0.09 & 0.25$\pm$0.05 &  & 0.24 & 0.28 &\\
 $A_2^{B K^*}(0)$  & 0.32 & 0.258$\pm$0.035 & 0.26$\pm$0.08 & 0.30$\pm$0.06 & 0.40$\pm$0.03 & 0.32 & &\\
 $A_2^{B_s\phi}(0)$  & 0.28 &  0.234$\pm$0.028 & & & & & &\\
 $T_1^{B \rho}(0)$ & 0.25 & 0.268$\pm$0.021 & 0.28$\pm$0.09 & 0.26$\pm$0.05 & &
 0.27 & &\\
 $T_1^{B K^*}(0)$  & 0.33 & 0.332$\pm$0.037 & 0.33$\pm$0.10 & 0.30$\pm$0.06 & 0.19$\pm$0.03 & 0.39 & &\\
 $T_1^{B_s\phi}(0)$  & 0.28 & 0.349$\pm$0.033  & & & & & &\\
\hline\hline
\end{tabular}
\end{center}
\end{table}


\begin{figure}[ht]
\begin{center}
\hspace*{-0.5cm}
\begin{tabular}{lcr}
\includegraphics[width=0.40\textwidth]{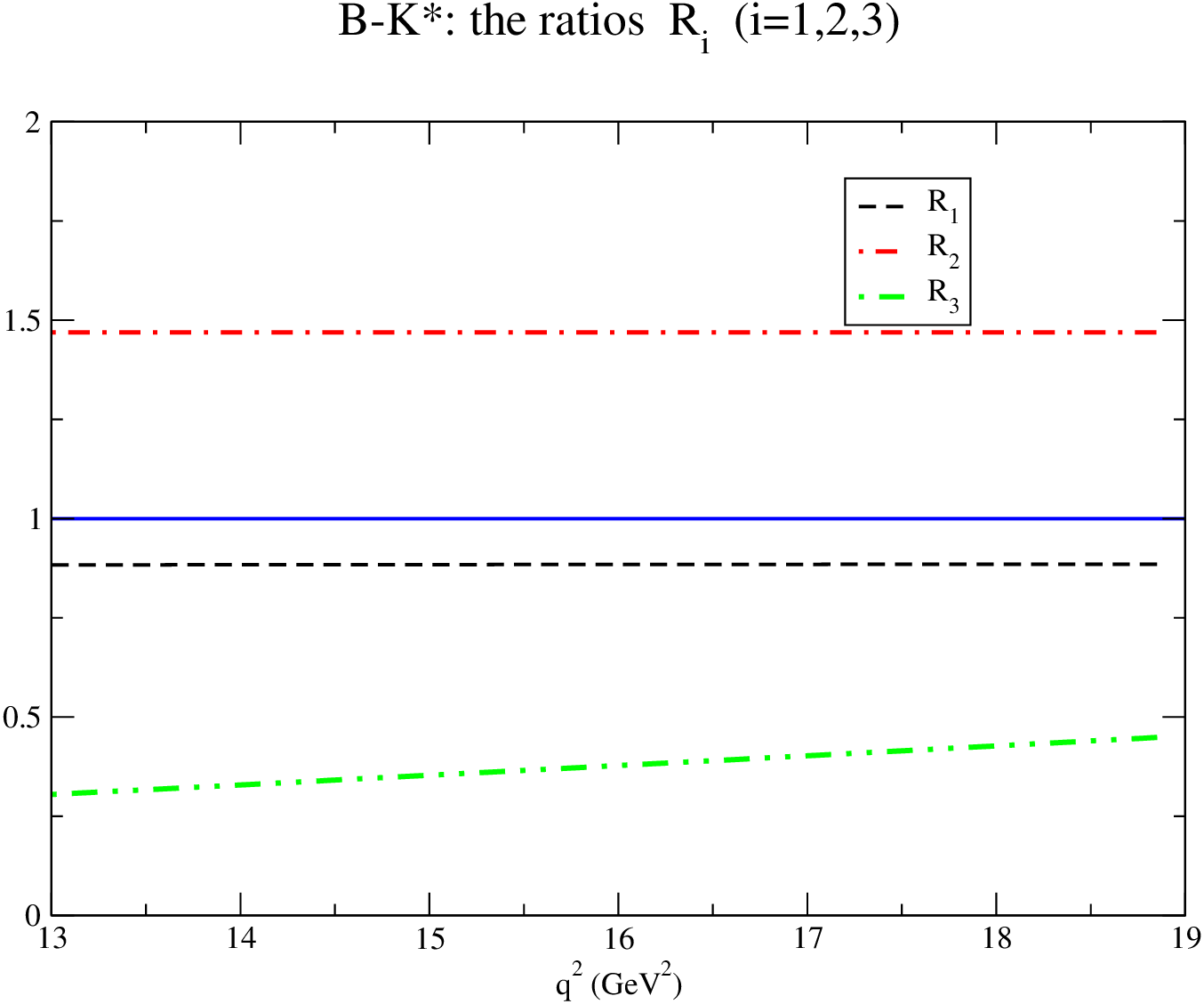} & \hspace{0.6truecm} &
\includegraphics[width=0.40\textwidth]{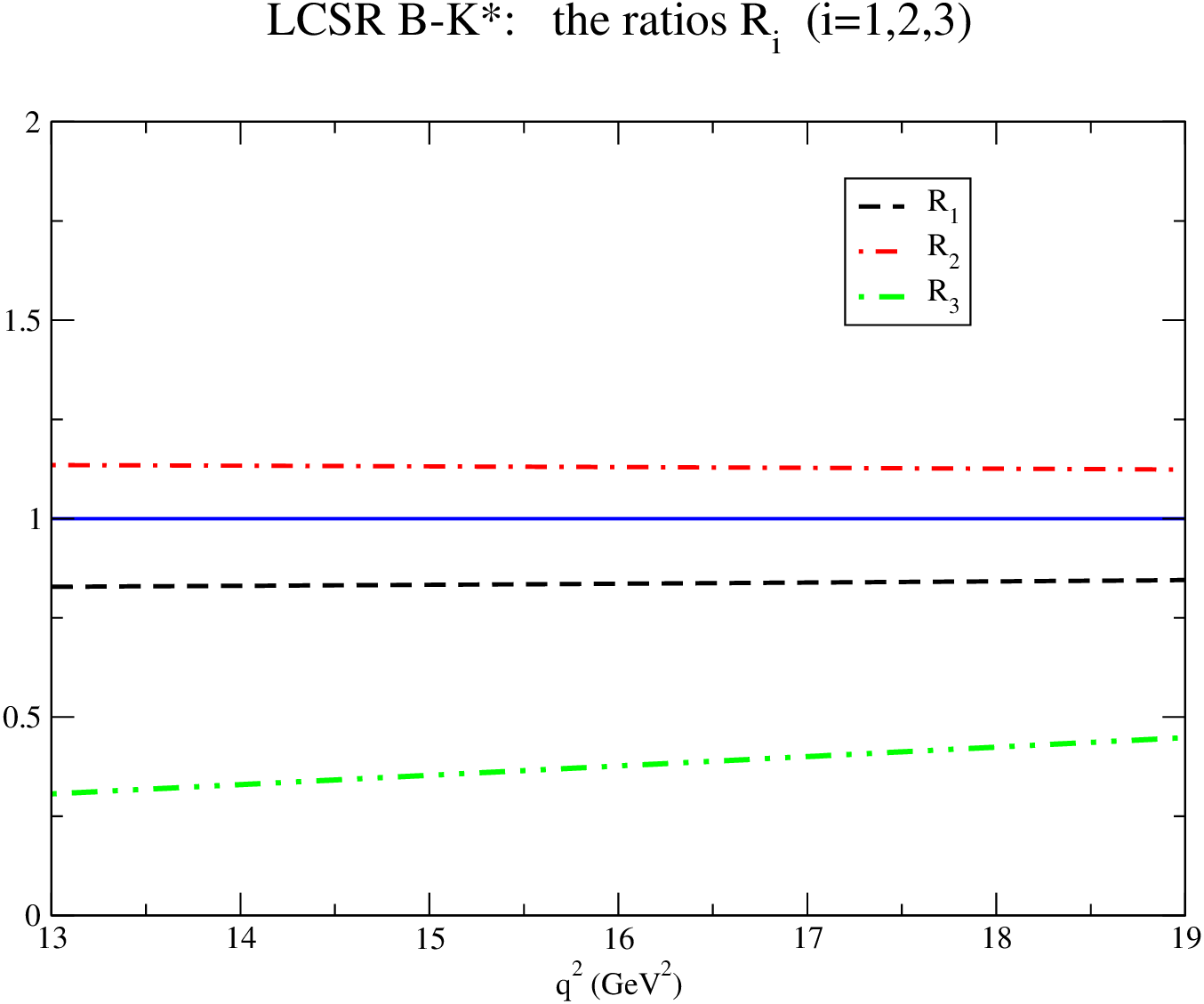}
\end{tabular}
\end{center}
\caption{\label{fig:ff_ratio} Our results for the ratios 
of the form factors appearing 
in Eq.\,(\protect\ref{eq:ff_ratio}) 
 for  $B-K^\ast-$transition.} 
\end{figure}

As was suggested in Ref.~\cite{Hiller}, one can check
how well the quark-model form factors satisfy the three low recoil relations
derived in Ref.\,\cite{Grinstein:2004vb} involving the pairs of form factors 
$(T_{1},V)$, $(T_{2},A_{1})$, and $(T_{3},A_{2})$. 
In Fig.~\ref{fig:ff_ratio}, we plot the ratios 
\be
R_1 = \frac{T_1(q^2)}{V(q^2)}, \qquad
R_2 = \frac{T_2(q^2)}{A_1(q^2)}, \qquad
R_3 = \frac{q^2}{m^2_B}\frac{T_3(q^2)}{A_2(q^2)}.
\label{eq:ff_ratio}
\en
which in the heavy-quark symmetry limit and at low recoil, should all be of 
the order
$1-(2\alpha_s/(3\pi)\ln\left(\mu/m_b\right)$, i.e.
close to one. Figure~\ref{fig:ff_ratio} shows that, similar to the 
extrapolated
LCSR form factors, the covariant quark model form factors satisfy the 
low-recoil heavy-quark symmetry relations 
reasonably well for $R_1$ and $R_2$ but not for $R_3$. Note that the $q^{2}$
scale has changed in  Fig.~\ref{fig:ff_ratio}.

\begin{figure}[ht]
\begin{center}
\includegraphics[width=0.40\textwidth]{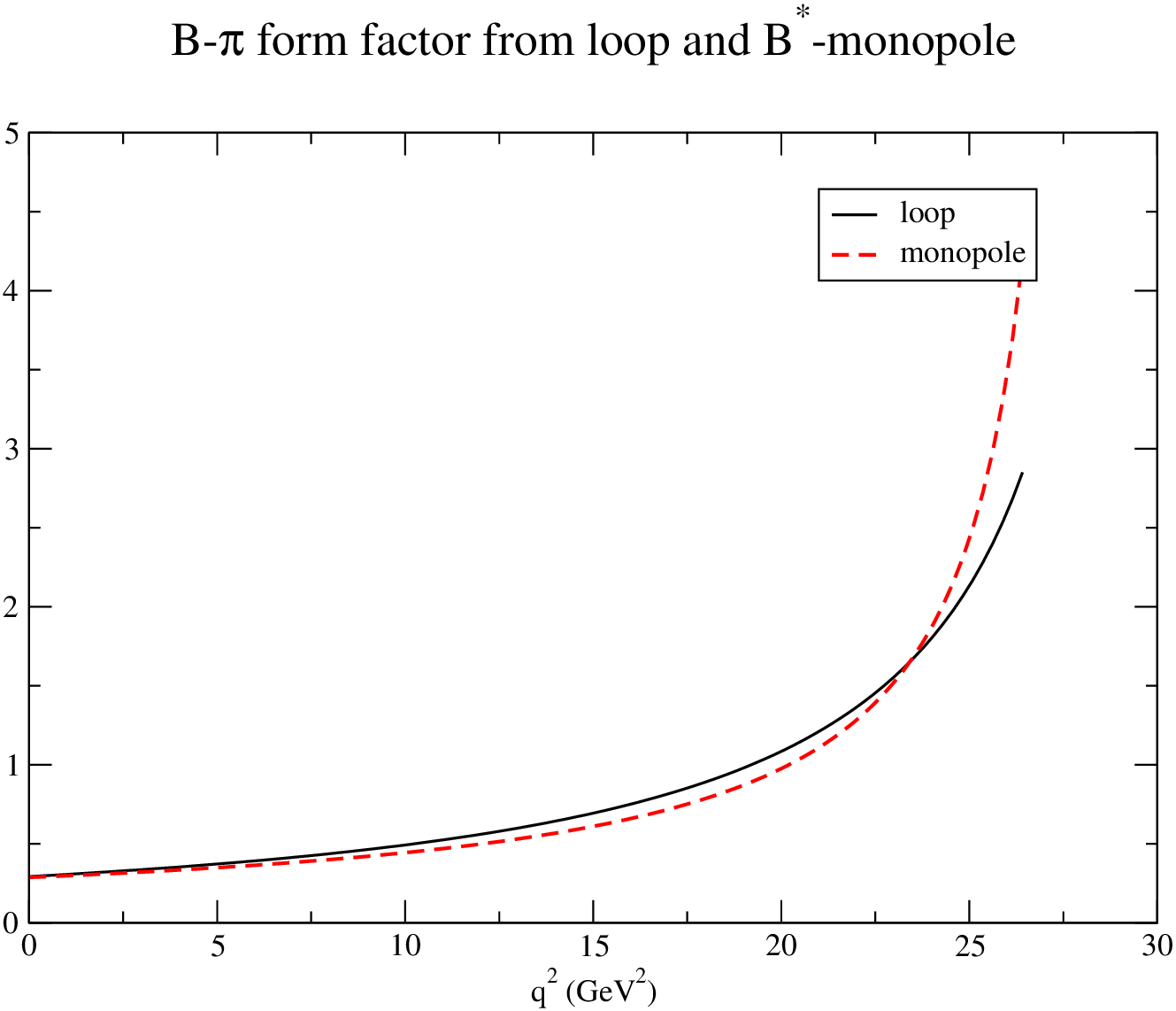} 
\end{center}
\caption{\label{fig:VDM} Comparison of the 
$B-\pi-$ form factors obtained
from the covariant quark model and 
from a VDM-monopole ansatz.
}
\end{figure}

It is interesting to compare the $q^{2}$ behavior of the $B-\pi$ transition
form factors calculated from the three-point one--loop diagram with the 
$q^{2}$ behavior of the vector-dominance model (VDM). For example,
in a monopole ansatz for the form factor $F_+^{B\pi}(q^2)$ one would have the 
VDM $q^{2}$ behavior 
\[
F_{\rm VDM}^{B\pi}(q^2) = \frac{F_+^{B\pi}(0)}{m^2_{B^\ast}-q^2}.
\label{eq:VDM}
\] 
where the pole mass is given by the mass of the lowest-lying vector meson 
state $B^{\ast}$. The two curves are plotted in Fig.~\ref{fig:VDM}. One 
observes a strong rise of the VDM form factor towards the
larger $q^{2}$ values close to the position of the $B^{\ast}$ pole. A similar
rise is observed for the quark-model form factor. It is quite intriguing
and gratifying that the quark-model form factor is able to emulate
the pole--type behavior of the VDM form factor including even the correct 
scale $m_{B}^{\ast}$ of the pole--type enhancement.

\section{Nonleptonic $B_{s}$ Decays}
\label{sec:nonlep}
As a second application we consider the two--body nonleptonic $B_{s}$
decays $B_{s} \to D_{s}^{(\ast)-} D_{s}^{(\ast)+}$
and $B_{s} \to \Jpsi\phi$, which have recently attracted some interest.
The modes  
$D_s^- D_s^+,$  $D_s^{\ast\,-} D_s^{+}+D_s^- D_s^{\ast\,+}$, and
$D_s^{\ast\,-} D_s^{\ast\,+}$ give the largest contribution
to $\Delta\Gamma$ for the $B_s-\bar B_s$ system.
The mode $\Jpsi\phi$ is color--suppressed 
but it is interesting for the search of possible CP-violating 
new physics effects in $B_s-\bar B_s$ mixing.

It is convenient to express all physical observables
in terms of helicity form factors $H_m$. This will result in very compact 
rate expressions. Furthermore, in the case of the two--vector meson decays
$P\to VV$, the helicity representation is quite convenient since one can
then easily calculate the helicity composition of the rate 
$\Gamma_{L},\Gamma_{-},\Gamma_{+}$.

The helicity form factors $H_m$ can be expressed in terms of
the invariant form factors of Ref.\,\cite{Khodjamirian:2006st} in the following 
way \cite{IKS}:

\vspace{0.5cm}
\noindent
(a) Spin $S=0$:

\begin{eqnarray}
H_t &=& \frac{1}{\sqrt{q^2}}
\left\{(m_1^2-m_2^2)\, F_+ + q^2\, F_- \right\}\,,
\nn
\label{helS0b}\\
H_0 &=& \frac{2\,m_1\,|{\bf p_2}|}{\sqrt{q^2}} \,F_+ \,.
\nonumber
\end{eqnarray}

\vspace{0.5cm}
\noindent
(b) Spin $S=1$:

\bea
H_t &=&
\frac{1}{m_1+m_2}\frac{m_1\,|{\bf p_2}|}{m_2\sqrt{q^2}}
\left\{ (m_1^2-m_2^2)\,(A_+ - A_0)+q^2 A_- \right\},
\nn
H_\pm &=&
\frac{1}{m_1+m_2}\left\{- (m_1^2-m_2^2)\, A_0
\pm 2\,m_1\,|{\bf p_2}|\, V \right\},
\label{helS1c}\\
H_0 &=&
\frac{1}{m_1+m_2}\frac{1}{2\,m_2\sqrt{q^2}}
\left\{-(m_1^2-m_2^2) \,(m_1^2-m_2^2-q^2)\, A_0
+4\,m_1^2\,|{\bf p_2}|^2\, A_+\right\},
\nonumber
\ena
where $|{\bf p_2}|=\lambda^{1/2}(m_1^2,m_2^2,q^2)/(2\,m_1)$
is the momentum of the outgoing particles
in the rest frame of the decaying particle.

The effective Hamiltonian describing the $B_s$ nonleptonic
decays is given by (see Ref.~\cite{Buchalla:1995vs})
\bea
{\mathcal H}_{\rm eff} &=&
-\frac{G_F}{\sqrt{2}}\,V_{cb} V^\dagger_{cs}\,\sum_{i=1}^6 C_i\,Q_i,
\nn
&&\nn
Q_1 &=& (\bar c_{a_1} b_{a_2})_{V-A} (\bar s_{a_2} c_{a_1})_{V-A}, 
\qquad 
Q_2  =  (\bar c_{a_1}\,b_{a_1})_{V-A}, (\bar s_{a_2}\,c_{a_2})_{V-A},
\nn
Q_3 &=& (\bar s_{a_1} b_{a_1})_{V-A} (\bar c_{a_2} c_{a_2})_{V-A}, 
\qquad 
Q_4  =  (\bar s_{a_1} b_{a_2})_{V-A} (\bar c_{a_2} c_{a_1})_{V-A},
\nn
Q_5 &=& (\bar s_{a_1} b_{a_1})_{V-A} (\bar c_{a_2} c_{a_2})_{V+A}, 
\qquad 
Q_6  =  (\bar s_{a_1} b_{a_2})_{V-A} (\bar c_{a_2} c_{a_1})_{V+A},
\label{eq:Hamilt}
\ena
where the subscript $V-A$ refers to the usual left--chiral current
$O^\mu_-=\gamma^\mu(1-\gamma^5)$ and  $V+A$  to the usual right--chiral one
$O^\mu_+=\gamma^\mu(1+\gamma^5)$. The  $a_i$ denote color indices. 
We calculate the nonleptonic $B_s$-decay widths by
using {\it naive} factorization. 
In this paper we consider the following nonleptonic decays of
the $B_s$-meson:
\[
B_s(p)\to 
 D_s^-(q_1) D_s^+(q_2),\; 
 D_s^-(q_1) D_s^{\ast\,+}(q_2,\epsilon_2),\; 
 D_s^{\ast\,-}(q_1,\epsilon_1)D_s^+(q_2),\;
 D_s^{\ast\,-}(q_1,\epsilon_1) D_s^{\ast\,+}(q_2,\epsilon_2),\;
\text{and}\; \Jpsi(q_1,\epsilon_1)\,\phi(q_2,\epsilon_2).
\]

The widths can be conveniently expressed in terms of the helicity form factors
and leptonic decay constants.
In the case of the color-allowed decays 
$B_{s} \to D_{s}^{(\ast)-} D_{s}^{(\ast)+}$ one has
\bea
\Gamma(B_s\to D_s^- D_s^+) &=&
\frac{G_F}{16\pi}\frac{|{\bf q_2}|}{m^2_{B_s}} [\lambda^{(s)}_c]^2 
\left( 
     C^{\,\rm eff}_{2}\,m_{D_s}\,f_{D_s}\,H_t^{B_sD_s}(m^2_{D_s})
 +2\,C^{\,\rm eff}_{6}\,f_{D_s}^{PS}\,F_S^{B_sD_s}(m^2_{D_s})
\right)^2\,,
\nn
\Gamma(B_s\to D_s^- D_s^{\ast\,+}) &=&
\frac{G_F}{16\pi}\frac{|{\bf q_{\,2}}|}{m^2_{B_s}} [\lambda^{(s)}_c]^2 
\left(     C^{\,\rm eff}_{2}\,m_{D_s}\,f_{D_s}\,H_t^{B_sD^\ast_s}(m^2_{D_s})
       +2\,C^{\,\rm eff}_{6}\,\frac{m_{B_s}|{\bf q_{\,2}}|}{m_{D_s^\ast}}
  f_{D_s}^{PS}\,F_{PS}^{B_sD_s^\ast}(m^2_{D_s})
\right)^2\,,
\nn
\Gamma(B_s\to D_s^{\ast\,-} D_s^{+}) &=&
\frac{G_F}{16\pi}\frac{|{\bf q_{\,2}}|}{m^2_{B_s}} [\lambda^{(s)}_c]^2  
\left(     C^{\,\rm eff}_{2}\,m_{D^\ast_s}\,f_{D^\ast_s}\,H_0^{B_sD_s}(m^2_{D^\ast_s})
\right)^2\,,
\nn
\Gamma(B_s\to D_s^{\ast\,-} D_s^{\ast\,+}) &=&
\frac{G_F}{16\pi}\frac{|{\bf q_{\,2}}|}{m^2_{B_s}} [\lambda^{(s)}_c]^2  
\left(     C^{\,\rm eff}_{2}\,m_{D^\ast_s}\,f_{D^\ast_s} \right)^2\,
\sum_{i=0,\pm}\left(H_i^{B_sD^\ast_s}(m^2_{D^\ast_s})\right)^2.
\label{eq:BsDD}
\ena
Here, $ \lambda^{(s)}_c=\frac{G_F}{\sqrt{2}}\,|V_{cb}V^\dagger_{cs}|$.
The Wilson coefficients appear in the combinations
$ C^{\,\rm eff}_{2}=C_2+\xi\, C_1+C_4+\xi\, C_3 $ and
$ C^{\,\rm eff}_{6}=C_6+\xi\, C_5.$ where terms multiplied by the color 
factor $\xi=1/N_c$
will be dropped in the numerical calculations according to the 
$1/N_c-$expansion. The annihilation channels that also contribute to the 
above color-allowed decays will be neglected
since they are color and form factor suppressed.

The width of the color-suppressed $B_s\to\Jpsi\,\phi$ decay 
is written as
\bea
\Gamma(B_s\to\Jpsi\,\phi ) &=&
\frac{G_F}{16\pi}\frac{|{\bf q_{\,2}}|}{m^2_{B_s}}  [\lambda^{(s)}_c]^2 
\left(C^{\,\rm eff}_1+ C^{\,\rm eff}_5\right)^2 
\left( m_{\Jpsi}\,f_{\Jpsi} \right)^2\,
\sum_{i=0,\pm}\left(H_i^{B_s \Jpsi}(m^2_{\Jpsi})\right)^2,
\label{eq:BsJpsiPhi}
\ena
where we have combined the Wilson coefficients into
$ C^{\,\rm eff}_{1}=C_1+\xi\, C_2+C_3+\xi\, C_4 $ and
$ C^{\,\rm eff}_{5}=C_5+\xi\, C_6.$.

For the Cabibbo-Kobayashi-Maskawa-matrix elements, we use the values from Ref.\,\cite{PDG}:
\be
\def\arraystretch{2}
\begin{array}{ccccccc}
   |V_{ud}|    &    |V_{us}|     & |V_{ub}|       & |V_{cd}| & |V_{cs}| & |V_{cb}| \\
\hline
\ \ 0.974 \ \ & \ \ 0.225 \ \  & \ \ 0.00389\ \ & \ \ 0.230 \ \  & 
\ \ 0.975 \ \  & \ \ 0.0406 \ \ \\
\end{array}
\label{eq:CKM}
\en

For the values of the Wilson coefficients we take
\cite{Altmannshofer:2008dz}
\be
\def\arraystretch{2}
\begin{array}{ccccccc}
   C_1    &    C_2  & C_3   & C_4 & C_5 & C_6 \\
\hline
\ \ -0.257 \ \ & \ \ 1.009 \ \  & \ \ - 0.005\ \ & \ \ -0.078 \ \  & 
\ \ 0.000 \ \  & \ \ 0.001 \ \ \\
\end{array}
\label{eq:Wilson}
\en
evaluated to next-to-next-to leading logarithmic accuracy 
in the $\overline{MS}$ (NDR) renormalization scheme at the scale
$\mu=4.8$~GeV \cite{Bobeth:1999mk}. 

We also need the values of the $B_s-\phi$ transition form factors evaluated
at $q^2=m^2_{\Jpsi}$. They are given in Table~\ref{tab:ff-BsPhi} 
where we compare our results with corresponding results of 
Ref.~\cite{Faller:2008gt}. 
The agreement for the form factors $A_1(m^2_{\Jpsi})$ and 
$A_2(m^2_{\Jpsi})$ is satisfactory. Our value for the form factor
$V(m^2_{\Jpsi})$ is somewhat smaller than the one found 
in Ref.~\cite{Faller:2008gt}. 

\begin{table}[ht]
\begin{center}
\def\arraystretch{1.5}
     \caption{The relevant $B_s-\phi$ form factors at $q^2=m^2_{\Jpsi}$
calculated in this work. For comparison, we give the results of 
Ref.~\cite{Faller:2008gt}.  }
\label{tab:ff-BsPhi}
\begin{tabular}{lcc}
\hline\hline
  & This work & Ref.~\cite{Faller:2008gt} \\
\hline
$A_1(m^2_{\Jpsi})$ & 0.37 & 0.42$\pm$0.06 \\
$A_2(m^2_{\Jpsi})$ & 0.48 & 0.38$\pm$0.06 \\
$V(m^2_{\Jpsi})$  & 0.56  & 0.82$\pm$0.12 \\
\hline\hline
\end{tabular}
\end{center}
\end{table}
In Table~\ref{tab:nonlep-widths} we give our results for the 
branching ratios. One can see that there is good
agreement with the available experimental data.

\begin{table}[ht]
\begin{center}
\def\arraystretch{1.5}
     \caption{Branching ratios ($\%$) of the $B_s$ nonleptonic decays
 calculated in our approach.}
\label{tab:nonlep-widths}
\vspace*{0.2cm}
\begin{tabular}{lcc}
\hline\hline 
Process &\,\,\, This work\,\,\, & Exp.\,data~\cite{PDG}  \\
\hline
$B_s\to D_s^- D_s^+ $     & 1.65 & $1.04^{+0.29}_{-0.26}$ \\
$B_s\to D_s^- D_s^{\ast\,+}+ D_s^{\ast\,-} D_s^{+}$  & 2.40 & $2.8 \pm 1.0 $ \\
$B_s\to D_s^{\ast\,-} D_s^{\ast\,+} $ & 3.18 & $3.1 \pm 1.4$\\
$B_s\to \Jpsi\phi$ & 0.16  & \,\,$0.14 \pm 0.05$\,\, \\
\hline\hline
\end{tabular}
\end{center}
\end{table}

We finally give our results on the helicity fractions in the two decays
$B_s\to D_s^{\ast\,-} D_s^{\ast\,+}$ and $B_s\to \Jpsi\phi$.
The helicity fractions of the nonleptonic $B_{s} \to VV$ rates are defined as
\be
\label{eq:helfrac}
\hat{\Gamma}_{L} = \frac{|H_0|^2}{|H_0|^2+|H_+|^2+|H_-|^2}, \qquad
\hat{\Gamma}_{\pm} =\frac{|H_\pm|^2}{|H_0|^2+|H_+|^2+|H_-|^2}, \qquad
\hat{\Gamma}_{\perp} = \frac12\frac{|H_+-H_-|^2}{|H_0|^2+|H_+|^2+|H_-|^2}.
\en
Note that we have normalized the partial helicity
rates to the total rate such that one has 
$(\hat{\Gamma}_{L}+\hat{\Gamma}_{-}+\hat{\Gamma}_{+})=1$.
For  $B_s\to D_s^{\ast\,-} D_s^{\ast\,+}$ we find 
$(\hat{\Gamma}_{L},\hat{\Gamma}_{-},\hat{\Gamma}_{+})$
=$(0.549,0.366,0.0847)$ and for $B_s\to \Jpsi\phi$ we find
$(0.420,0.552,0.0272)$.
The hierarchy of partial helicity rates 
$\hat{\Gamma}_{L}>\hat{\Gamma}_{-}>\hat{\Gamma}_{+}$ seen in the decay
$B_s\to D_s^{\ast\,-} D_s^{\ast\,+}$ is expected for
tree-level-dominated nonleptonic decays using simple on--shell quark model 
arguments. One finds that, at the leading order of $m_1=m_{B_s}$,
the partial rate $\Gamma_{-}$ is helicity-suppressed by the factor 
$4 q^2/m_1^2$ with $q^2=m^2_{D_s^{\ast\,+}}$
and the partial rate $\Gamma_{+}$ is further chirality suppressed by the 
factor $m_2^2/m_1^2$ with $m_2=m_{D_s^{\ast\,+}}$ in addition to
the helicity suppression~\cite{DESY-79-60,DESY 78/51,hep-ph/0612290}.
Using the qualitative suppression factors one finds 
$(0.583,0.361,0.056)$ for the helicity fractions in the decay 
$B_s\to D_s^{\ast\,-} D_s^{\ast\,+}$ which is 
remarkably close to the results of the full calculation. For the process 
$B_s\to \Jpsi\phi$ with a larger $q^2$--value of $q^2=m^2_{\Jpsi}$ the 
helicity suppression is no longer in effect since now $4 q^2/m_1^2=1.332$.
One now obtains  $(0.420,0.560,0.020)$ for the helicity fractions which again 
is remarkably close to the results of the full calculation.
One has an inversion of the 
hierarchy for the longitudinal and transverse--minus rates for 
$B_s\to \Jpsi\phi$ in as much
as one now has $\hat{\Gamma}_{L}<\hat{\Gamma}_{-}$. 
Experimental numbers
on the partial helicity rates exist only for the decay $B_s\to \Jpsi\phi$ 
given by $\hat{\Gamma}_{L}=0.541 \pm 0.017$ and 
$\hat{\Gamma}_{\perp}=0.241 \pm 0.023$~\cite{PDG}. Our calculated
longitudinal rate can be seen to be off by several standard deviations. In 
order to be able to compare with the experimental transverse rate 
$\hat{\Gamma}_{\perp}$ one needs to use 
$\Gamma_{\perp}\propto |A_{\perp}|^{2}=|H_{+}-H_{-}|^{2}/2$. 
For  $B_s\to D_s^{\ast\,-} D_s^{\ast\,+}$ we find $\hat{\Gamma}_{\perp}=0.0493$
and for $B_s\to \Jpsi\phi$ we predict $\hat{\Gamma}_{\perp}=0.167$. Again we
are off the experimental result by several standard deviations.

\section{Summary}
\label{sec:summary}
We have given a brief sketch of the
theoretical framework underlying the covariant quark model, including a 
discussion of how infrared confinement is incorporated in the model. 
We have discussed in some detail how the
model parameters of the covariant quark model are determined
through a least-squares fit to experimental/theoretical data on the leptonic
decay constants and eight fundamental mesonic one-- and two--photon decays. 
Once the model parameters of the covariant quark model are fixed the model 
can be used to obtain parameter-free predictions for any transition process 
involving light or heavy mesons. 

In the present paper, we have calculated the transition form factors 
of the heavy $B$ and $B_{s}$ mesons to light pseudoscalar and vector mesons,
which are needed as ingredients for the calculation of the semileptonic,
nonleptonic, and rare decays of the $B$ and $B_{s}$ mesons. Our form factor
results hold in the full kinematical range of momentum transfer. We have 
provided a detailed discussion of how the covariant-quark-model form factors 
compare with the corresponding LCSR form factors. 

 We have finally made use of the calculated form factors to
calculate the nonleptonic decays $B_{s} \to D_{s}^{(\ast)-} D_{s}^{(\ast)+}$
and $B_{s} \to \Jpsi\phi$, which have been widely discussed recently in the
context of $B_s-\bar B_s$--mixing and CP violation. We have also presented 
results on the helicity composition for the decays $B_{s} \to VV$. 
Further application of our form factor results are envisaged, such as the
calculation of the penguin--dominated decay 
$B_{s}\to K^{(\ast)}\overline{K}^{(\ast)}$.

\begin{acknowledgments}

This work was supported by
the DFG Grant No. KO 1069/13-1,
the Heisenberg-Landau program,
Russian Fund of Basic Research Grant No. 10-02-00368-a,
\mbox{FONDECYT} projects 1100582 and
Centro-Cient\'\i fico-Tecnol\'{o}gico de Valpara\'\i so PBCT ACT-028.

\end{acknowledgments}


\clearpage
\begin{thebibliography}{99}

\bibitem{Stone} S~Stone, "Heavy Flavor Physics",[arXiv:1109.3361 [hep-ph]].
To appear in Proceedings of the DPF-2011 Conference, 
Providence, RI, August 8-13, 2011.

\bibitem{leitner11}
    B. El-Bennich, J. P. B. C. de Melo, O. Leitner, B. Loiseau, J.-P.
Dedonder, [arXiv:1111.6955 [hep-ph]], in Erice School on Nuclear Physics
2011: From Quarks and Gluons to Hadrons and Nuclei
(unpublished). 

\bibitem{Faller:2008gt}
  S.~Faller, R.~Fleischer, T.~Mannel,
  Phys.\ Rev.\  {\bf D79}, 014005 (2009).
  [arXiv:0810.4248 [hep-ph]].


\bibitem{Buchalla:1995vs}
  G.~Buchalla, A.~J.~Buras, M.~E.~Lautenbacher,
  Rev.\ Mod.\ Phys.\  {\bf 68}, 1125-1144 (1996).
  [hep-ph/9512380].





\bibitem{Beneke:1999br} 
  M.~Beneke, G.~Buchalla, M.~Neubert and C.~T.~Sachrajda,
  Phys.\ Rev.\ Lett.\  {\bf 83}, 1914 (1999)
  [hep-ph/9905312];
%
  M.~Beneke, M.~Neubert,
  Nucl.\ Phys.\  {\bf B675}, 333-415 (2003).
  [hep-ph/0308039].


\bibitem{Feldmann:2008sk}
  T.~Feldmann,
  ``Soft-Collinear Effective Theory: Recent Results and Applications,''
  PoS {\bf CONFINEMENT8}, 007 (2008).
  [arXiv:0811.4590 [hep-ph]].


\bibitem{Bauer:2000yr}
  C.~W.~Bauer, S.~Fleming, D.~Pirjol, I.~W.~Stewart,
  Phys.\ Rev.\  {\bf D63}, 114020 (2001).
  [hep-ph/0011336].


\bibitem{Ball}
  P.~Ball and R.~Zwicky,
  Phys.\ Rev.\  D {\bf 71}, 014029 (2005) [arXiv:hep-ph/0412079].
  
 
\bibitem{Khodjamirian:2006st}
  A.~Khodjamirian, T.~Mannel, N.~Offen,
  Phys.\ Rev.\  {\bf D75}, 054013 (2007). [hep-ph/0611193].



\bibitem{Grinstein:2004vb}
  B.~Grinstein, D.~Pirjol,
  Phys.\ Rev.\  {\bf D70}, 114005 (2004).
  [hep-ph/0404250].


\bibitem{Hiller} 
  C.~Bobeth, G.~Hiller, D.~van Dyk,
  JHEP {\bf 1007}, 098 (2010)
  [arXiv:1006.5013 [hep-ph]];
%
  JHEP {\bf 1107}, 067 (2011).
  [arXiv:1105.0376 [hep-ph]].



\bibitem{Ivanov:2007cw}
  M.~A.~Ivanov, J.~G.~K\"orner, S.~G.~Kovalenko, C.~D.~Roberts,
  Phys.\ Rev.\  {\bf D76}, 034018 (2007).
  [nucl-th/0703094].

\bibitem{Melikhov}
  D.~Melikhov, N.~Nikitin and S.~Simula,
  Phys.\ Rev.\  D {\bf 57}, 6814 (1998), [arXiv:hep-ph/9711362];
%
  D.~Melikhov,
  Eur.\ Phys.\ J.\ direct C {\bf 4}, 1 (2002), [arXiv:hep-ph/0110087].

\bibitem{Ebert:2006nz}
  D.~Ebert, R.~N.~Faustov, V.~O.~Galkin,
  Phys.\ Rev.\  {\bf D75}, 074008 (2007).
  [hep-ph/0611307].

\bibitem{Ladisa:1999cu}
  M.~Ladisa, G.~Nardulli, P.~Santorelli,
  Phys.\ Lett.\  {\bf B455}, 283-290 (1999).
  [hep-ph/9903206].
  
\bibitem{hep-ph/9809372} 
  P.~Colangelo, F.~De Fazio, M.~Ladisa, G.~Nardulli, P.~Santorelli and A.~Tricarico,
  Eur.\ Phys.\ J.\ C\ {\bf 8}, 81  (1999)
  [hep-ph/9809372].

\bibitem{Colangelo:1995jv}
  P.~Colangelo, F.~De Fazio, P.~Santorelli, E.~Scrimieri,
  Phys.\ Rev.\  {\bf D53}, 3672-3686 (1996).
  [hep-ph/9510403];
%
%
  P.~Colangelo, P.~Santorelli,
  Phys.\ Lett.\  {\bf B327}, 123-128 (1994).
  [hep-ph/9312258].




\bibitem{Faessler:2002ut}
  A.~Faessler, T.~Gutsche, M.~A.~Ivanov, J.~G.~K\"orner and V.~E.~Lyubovitskij,
  Eur.\ Phys.\ J.\ directC {\bf 4}, 18 (2002)
  [arXiv:hep-ph/0205287].

\bibitem{IKS}
  M.~A.~Ivanov, J.~G.~K\"{o}rner and P.~Santorelli,
  Phys.\ Rev.\ D {\bf 63}, 074010 (2001)
  [arXiv:hep-ph/0007169];
%
  Phys.\ Rev.\ D {\bf 71}, 094006 (2005)
  [arXiv:hep-ph/0501051];
  Phys.\ Rev.\  {\bf D73}, 054024 (2006)
  [hep-ph/0602050].

\bibitem{Ivanov:2002un}
  M.~A.~Ivanov, J.~G.~K\"{o}rner and O.~N.~Pakhomova,
  Phys.\ Lett.\ B {\bf 555}, 189 (2003)
  [arXiv:hep-ph/0212291].

\bibitem{Z=0} A.~Salam,
  Nuovo Cim.\  {\bf 25}, 224 (1962);
S.~Weinberg,
  Phys.\ Rev.\  {\bf 130}, 776 (1963);

\bibitem{Z=0_2} For a review, see: K.~Hayashi, M.~Hirayama, T.~Muta, N.~Seto 
              and T.~Shirafuji, Fort.\ der Phys.\ {\bf 15}, 625 (1967).
  

\bibitem{Branz:2009cd}
  T.~Branz, A.~Faessler, T.~Gutsche, M.~A.~Ivanov, J.~G.~K\"orner, 
  V.~E.~Lyubovitskij,
  Phys.\ Rev.\  {\bf D81}, 034010 (2010).
  [arXiv:0912.3710 [hep-ph]].

\bibitem{Dubnicka:2010kz}
  S.~Dubnicka, A.~Z.~Dubnickova, M.~A.~Ivanov, J.~G.~K\"orner,
  Phys.\ Rev.\  {\bf D81}, 114007 (2010).
  [arXiv:1004.1291 [hep-ph]].

\bibitem{Dubnicka:2011mm}
  S.~Dubnicka, A.~Z.~Dubnickova, M.~A.~Ivanov, J.~G.~K\"orner, P.~Santorelli, 
  G.~G.~Saidullaeva,
  Phys.\ Rev.\  {\bf D84}, 014006 (2011).
  [arXiv:1104.3974 [hep-ph]].




\bibitem{QCM}
G.~V.~Efimov and M.~A.~Ivanov,
``The Quark Confinement Model Of Hadrons,''
{\it  Bristol, UK: IOP (1993) 177 p};
Int.\ J.\ Mod.\ Phys.\ A {\bf 4}, 2031 (1989).
%
\bibitem{RCQM}
M.~A.~Ivanov, M.~P.~Locher and V.~E.~Lyubovitskij,
Few Body Syst.\  {\bf 21}, 131 (1996)
[arXiv:hep-ph/9602372];
M.~A.~Ivanov and V.~E.~Lyubovitskij,
Phys.\ Lett.\ B {\bf 408}, 435 (1997)
[arXiv:hep-ph/9705423].


\bibitem{Vermaseren:2000nd}
   J.~A.~M.~Vermaseren,
   Nucl.\ Phys.\ Proc.\ Suppl.\  {\bf 183}, 19 (2008)
   [arXiv:0806.4080 [hep-ph]];
   arXiv:math-ph/0010025.
   




\bibitem{PDG}
  K.~Nakamura {\it et al.} [ Particle Data Group Collaboration ],
  J.\ Phys.\ G {\bf G37 } (2010)  075021 (see also the 2011 update).



  
\bibitem{RosnerStone}
  J.~L.~Rosner, S.~Stone,
[arXiv:1002.1655 [hep-ex]].
   
\bibitem{LatticePRD81}
  J.~Laiho, E.~Lunghi, R.~S.~Van de Water,
  Phys.\ Rev.\  {\bf D81}, 034503 (2010).
  [arXiv:0910.2928 [hep-ph]].   
 
\bibitem{LatticeTWQCD}
 T.~-W.~Chiu {\it et al.} [ TWQCD Collaboration ],
 Phys.\ Lett.\  {\bf B651}, 171-176 (2007).
 [arXiv:0705.2797 [hep-lat]].
   
\bibitem{Lubicz}
  D.~Becirevic, P.~Boucaud, J.~P.~Leroy, V.~Lubicz, G.~Martinelli, 
  F.~Mescia, F.~Rapuano,
  Phys.\ Rev.\  {\bf D60}, 074501 (1999).
  [hep-lat/9811003].


\bibitem{Altmannshofer:2008dz}
  W.~Altmannshofer, P.~Ball, A.~Bharucha, A.~J.~Buras, D.~M.~Straub, M.~Wick,
  JHEP {\bf 0901}, 019 (2009).
  [arXiv:0811.1214 [hep-ph]].

\bibitem{Bobeth:1999mk}
  C.~Bobeth, M.~Misiak, J.~Urban,
  Nucl.\ Phys.\  {\bf B574}, 291-330 (2000).
  [hep-ph/9910220].

\bibitem{DESY-79-60}
  J.~G.~K\"orner and G.~R.~Goldstein,
  Phys.\ Lett.\ B\ {\bf 89} (1979) 105.

\bibitem{DESY 78/51}
  A.~Ali, J.~G.~K\"orner, G.~Kramer and J.~Willrodt,
  Z.\ Phys.\ C\ {\bf 1} (1979) 269.

\bibitem{hep-ph/0612290}
  M.~Beneke, J.~Rohrer and D.~Yang,
  Nucl.\ Phys.\ B\ {\bf 774} (2007) 64
  [hep-ph/0612290].
\end{thebibliography}
\end{document}